\documentclass[sigconf,nonacm,screen]{acmart}

\copyrightyear{2022}
\acmYear{2022}
\setcopyright{rightsretained}
\acmConference[CCS '22]{Proceedings of the 2022 ACM SIGSAC
Conference on Computer and Communications Security}{November
7--11, 2022}{Los Angeles, CA, USA}
\acmBooktitle{Proceedings of the 2022 ACM SIGSAC Conference on
Computer and Communications Security (CCS '22), November 7--11,
2022, Los Angeles, CA, USA}
\acmDOI{10.1145/3548606.3559341}
\acmISBN{978-1-4503-9450-5/22/11}

\usepackage{etoolbox}

\makeatletter

\patchcmd{\maketitle}{\@copyrightpermission}{

   \begin{minipage}{0.3\columnwidth}

     \href{https://creativecommons.org/licenses/by/4.0/}{\includegraphics[width=0.80\textwidth]{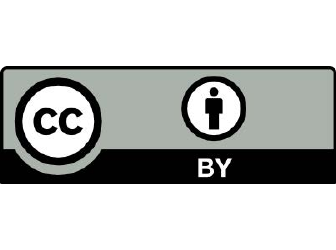}}

   \end{minipage}\hfill

   \begin{minipage}{0.7\columnwidth}

     \href{https://creativecommons.org/licenses/by/4.0/}{This work is licensed under a Creative Commons Attribution International 4.0 License.}

   \end{minipage}

   \vspace{5pt}

}{}{}

\makeatother

\usepackage{booktabs} 
\usepackage[ruled]{algorithm2e} 
\usepackage[tight]{subfigure}
\usepackage[utf8]{inputenc}
\usepackage{tikz}
\usepackage{threeparttable}
\usepackage{array}
\usepackage{geometry}
\usepackage[utf8]{inputenc}
\usepackage{graphicx} 
\usepackage{pgfplots}
\usepackage{caption}
\usepackage{graphicx}
\usepackage{bbm}
\usepackage{threeparttable}
\usepackage{soul}
\usepackage[misc]{ifsym}
\usepackage[tight]{subfigure}
\usepackage[capitalize]{cleveref}
\usepackage{rotating}
\usepackage{float}
\usepackage{longtable}
\usepackage{array}
\usepackage{amsmath}
\usepackage{tabularx}
\usepackage{ragged2e}
\usepackage{lipsum}
\usepackage{longtable}
\usepackage{enumitem}
\usepackage{setspace}
\usepackage[subtle,tracking=normal, wordspacing=normal,mathdisplays=tight]{savetrees} 

\newif\ifshowcomment
\showcommenttrue

\newcounter{todocounter}
\ifshowcomment
\newcommand{\luyao}[1]{\textcolor{red}{[luyao (camera ready revisions)] #1}}
\newcommand{\yuxuan}[1]{\textcolor{orange}{[Yuxuan] #1}}
\newcommand{\william}[1]{\textcolor{blue}{\footnotesize [William: #1]}}
\newcommand{\kartik}[1]{\textcolor{magenta}{\footnotesize [Kartik: #1]}}
\newcommand{\fanz}[1]{\stepcounter{todocounter}\textcolor{blue}{{\small [[(\thetodocounter) FZ: #1]]}}}
\newcommand{\yulin}[1]{\textcolor{red}{{\sf [Yulin: #1]}}}
\newcommand{\TODO}[1]{\textcolor{red}{{\sf [TODO: #1]}}}
\newcommand{\added}[1]{\textcolor{yellow}{[added(rebuttal)] #1}}

\else
\newcommand{\yuxuan}[1]{}
\newcommand{\william}[1]{}
\newcommand{\luyao}[1]{}
\newcommand{\kartik}[1]{}
\newcommand{\fanz}[1]{}
\newcommand{\yulin}[1]{}
\newcommand{\TODO}[1]{}
\newcommand{\added}[1]{}
\fi

\newcommand{\boldhead}[1]{\vspace{2pt}\noindent\textbf{#1.}}

\usepackage{xcolor}

\pgfplotsset{compat=1.17}

\begin{document}

\title{Empirical Analysis of EIP-1559: Transaction Fees, Waiting Times, and Consensus Security}

\author{Yulin Liu}
\affiliation{%
  \institution{SciEcon CIC}
  \country{United Kingdom}
}
\affiliation{%
  \institution{Bochsler Group}
  \country{Switzerland}
}
\authornotemark[3]

\author{Yuxuan Lu}
\affiliation{%
  \department{Center on Frontiers of Computing Studies}
  \institution{Peking University}
  \country{China}
}
\authornotemark[3]

\author{Kartik Nayak}
\affiliation{%
  \department{Department of Computer Science}
  \institution{Duke University}
  \country{United States}
}
\authornotemark[3]

\author[Fan Zhang*]{Fan Zhang}
\affiliation{%
  \department{Department of Computer Science}
  \institution{Yale University}
  \country{United States}
}
\authornotemark[1]
\authornotemark[2]
\authornotemark[3]

\author[Luyao Zhang*]{Luyao Zhang}
\affiliation{%
  \department{Data Science Research Center and Social Science Division}
  \institution{Duke Kunshan University}
  \country{China}
}

\authornote{Corresponding authors: \newline
Fan Zhang (email: f.zhang@yale.edu, address: 51 Prospect St, New Haven, CT 06520, United States) and Luyao Zhang (email: lz183@duke.edu, address: Duke Kunshan University, No.8 Duke Ave. Kunshan, Jiangsu 215316, China.) }
\authornote{The joint research was completed during the authors' Duke University and/or Duke Kunshan University (DKU) appointments.}
\authornote{The authors are listed in alphabetical order according to last names and, then, first names, and these authors contributed equally to this work.}
\authornote{Also with SciEcon CIC, 71-75 Shelton Street, Covent Garden, London, United Kingdom, WC2H 9JQ}

\author{Yinhong Zhao}
\affiliation{%
  \institution{Duke University}
  \country{United States}
}
\authornotemark[3]
\authornotemark[4]


\begin{abstract}
A transaction fee mechanism (TFM) is an essential component of a blockchain protocol. However, a systematic evaluation of the real-world impact of TFMs is still absent. Using rich data from the Ethereum blockchain, the mempool, and exchanges, we study the effect of EIP-1559, one of the earliest-deployed TFMs that depart from the traditional first-price auction paradigm. We conduct a rigorous and comprehensive empirical study to examine its causal effect on blockchain transaction fee dynamics, transaction waiting times, and consensus security. Our results show that EIP-1559 improves the user experience by mitigating intrablock differences in the gas price paid and reducing users' waiting times. However, EIP-1559 has only a small effect on gas fee levels and consensus security. In addition, we find that when Ether's price is more volatile, the waiting time is significantly higher. We also verify that a larger block size increases the presence of siblings. These findings suggest new directions for improving TFMs. 
\end{abstract}



\maketitle

\section{Introduction}
Computation and storage on public blockchains such as Bitcoin and Ethereum are scarce resources~\cite{buterin_2018_blockchain}. To allocate blockchain resources to users, a Transactions Fee Mechanism (TFM) must be employed. A TFM is an essential component of a blockchain protocol that can fundamentally affect the incentive compatibility, user experience, and security of a blockchain system~\cite{roughgarden_2020_transaction, ferreira_2021_dynamic,chung_2022_foundations, easley_2019_from, harvey_2021_defi}. Ethereum, for example, used to employ first-price auctions as the transaction fee mechanism~\cite{roughgarden_2020_transaction}. 

While many have proposed novel TFMs beyond simple first-price auctions~\cite{basu_2021_stablefees,ferreira_2021_dynamic,reijsbergen_2021_transaction,lavi_2019_redesigning,yao_2018_an}, there was no  real-world implementation until the Ethereum Improvement Proposal 1559 (EIP-1559)~\cite{ethereumimprovementproposals_2021_ethereum} on Ethereum, the second-largest blockchain network by market capitalization to date.

On August 5th, 2021, Ethereum activated a major upgrade named the London hard fork~\cite{ethereumfoundation_2021_london}, which implemented EIP-1559 together with several other EIPs and overhauled the Ethereum TFM. EIP-1559 introduced several novel elements while maintaining backward compatibility. Notably, for instance, it includes a base fee parameter that indicates the minimum gas price users need to pay in each block, which adjusts dynamically according to the gas used in the previous block. It also changes how users specify transaction fee bids. We defer our presentation of more details on EIP-1559 to~\cref{sec:background}.

To the best of our knowledge, EIP-1559 is not only the first major TFM change on Ethereum but also the first real attempt to depart from first-price auctions on any major blockchain. The impact of this upgrade is  profound. Multiple prior works have examined EIP-1559 from a theoretical point of view. Roughgarden~\cite{roughgarden_2020_transaction} gives a thorough game-theoretical evaluation of the EIP-1559 mechanism and points out its incentive compatibility for myopic miners. Reijsbergen et al.~\cite{reijsbergen_2021_transaction} observes the volatile gas usage after EIP-1559 and proposed modifications to mitigate this issue. 
The Ethereum community had analyzed the EIP informally~\cite{fellowshipofethereummagicians_2019_eip1559, buterin_2021_eip} and expected the upgrade to mitigate economic inefficiencies due to fee volatility, to prevent over-payment of transaction fees, and to lower transaction waiting times~\cite{ethereumimprovementproposals_2021_ethereum}. However, the real-world impact of a novel TFM such as EIP-1559 has not been systematically studied. 

We aim to close this gap with a comprehensive and rigorous empirical study. As a major and probably the only recent TFM reform, EIP-1559 presents a unique opportunity to study the causal effects of TFM changes on blockchain characteristics. While we focus on Ethereum, the insight we gain can generalize to other blockchains and future TFM reforms. We aim to answer three questions on the impact of this TFM reform.

\begin{itemize}[leftmargin=*]
    \item Does EIP-1559 affect the transaction fee dynamics? Existing theoretical studies predict easier fee estimation under the novel TFM because the Symmetric Ex-post Equilibrium (SEE) is easier to solve than the Bayesian Nash Equilibrium (BNE) in the previous first-price auction for bounded rational users.~\cite{roughgarden_2020_transaction,ferreira_2021_dynamic}.
    However, the rationality of users on Ethereum has yet to be tested. Thus, it is essential to verify the theoretical implications empirically. 
    \item Does EIP-1559 affect transaction waiting times? The Ethereum community expects the TFM to reduce transaction delays~\cite{buterin_2021_eip}, but it is unclear whether and how this happens.
    \item Does EIP-1559 affect consensus security? EIP-1559 introduces significant changes to the block size (in terms of gas used) and the incentive system of miners and users. The security implications are widely debated~\cite{eip_1559_spikes, fellowshipofethereummagicians_2019_eip1559, chung_2022_foundations}, but little real-world evidence is known. We aim to settle the arguments with empirical evidence.
\end{itemize}

\subsection*{Challenges and our approach}

To answer these questions, we collected rich data from the Ethereum blockchain, mempool\footnote{On Ethereum, mempool is where transactions stay after sent by users and before being added to a block by miners.} (for computing waiting time), and exchanges (e.g., intraday ETH prices)~\cite{googlecloudplatform_2019_big, bloomberglp_2019_xetusd}. Measurement of many blockchain characteristics is challenging. For example, measuring the waiting time of transactions requires accurate observations of the mempool.

We set up a distributed data collection system to monitor the mempool of Ethereum and capture the timestamps when each transaction is submitted to the mempool, thus obtaining a much more precise measure of transaction waiting time than the measures used in existing literature.

Empirically, it is difficult to separate the effect of EIP-1559 on blockchain characteristics from other confounding factors, such as price volatility, network instability, and the time trend. An empirical study aiming at unbiased estimates must control for these confounding factors. Thus, we adopt an event study~\cite{mackinlay_1997_event} and Regression Discontinuity Design (RDD)~\cite{lee_2010_regression, athey_2017_the} framework that enables the estimation of causal effects. By comparing observations of data on either side of the London hard fork, we estimate the local average treatment effect of EIP-1559.

\subsection*{Our Findings}

\boldhead{Transaction fees}
We observe that EIP-1559 did not lower the transaction fee level itself in our data period, but enabled easier fee estimation for users. 

Before EIP-1559, users paid the entirety of their bids, so they risked {\em overpaying} transaction fees if the network condition turned out to be less congested after they bid. With the new TFM, however, such risks are avoided, because users can set two parameters in their bids: a cap on the total fees that they will pay per gas (called the ``max fee per gas'') and a tip for the miner on top of the base fee (called the ``max priority fee per gas''). The actual fee paid is either the max fee per gas or the sum of the base fee and the max priority fee per gas, whichever is smaller. More details of EIP-1559 are provided below in~\cref{subsec:background:eip1559}.

This separation enables a simple yet optimal bidding strategy (dubbed the obvious optimal bid in~\cite{roughgarden_2020_transaction}) where users just set the max fee per gas to their intrinsic value for the transaction and set the max priority fee per gas to the marginal cost of miners. As we elaborate in~\cref{subsec:fee}, we observe that the bids that users submit after EIP-1559 are consistent with this obvious optimal bid. We also observe that users who adopt EIP-1559 bidding pay a lower fee than those who stick to legacy bidding. Both findings imply that fee estimation is easier with the new gas fee bidding style. Moreover, our regression discontinuity analysis in~\cref{subsec:IQR} indicates that the intrablock gas price variance, measured by the standardized inter-quartile range (IQR), becomes significantly lower as more users adopt EIP-1559 transactions. Therefore, the variance of intrablock gas prices decreases with EIP-1559, which also implies easier fee estimation and less overpayment for users. Our results thus imply that future mechanism designers should consider players' bounded rationality and design mechanisms easier for users to understand. 

\boldhead{Transaction waiting times.}
We observe that EIP-1559 lowers transaction waiting time, thus improving the user experience. 

We define the waiting time as the difference between the time when we first observe the transaction in the mempool and when the transaction is mined. 
The waiting time determines the latency of the commit. Moreover, when there are dependent transactions, users cannot submit new transactions until previous dependent transactions are successfully included in blocks or canceled. Thus, the delay has an opportunity cost associated with it. 

We find that the waiting time significantly declines after the London hard fork, possibly as a result of easier gas price bidding and variable-sized blocks. This benefits both the transactions that adopt the new bid style and the ones that still adopt legacy bidding. Thus, EIP-1559 has improved the waiting time for transactions even though not all users have adopted it. The reduction in waiting times might also be a consequence of the easier fee estimation under EIP-1559.The true value of bidding reveals the opportunity cost of time. With a more obvious optimal bidding strategy, users with more urgent needs bid higher to have their transactions included in the next available block. 

\boldhead{Consensus security}
EIP-1559 changes important consensus parameters such as the block size and the incentive of miners and users.
To understand its impact on consensus security, we identified three possible avenues through which the EIP might affect consensus:
\begin{itemize}[leftmargin=*]
    \item {\em Fork rate.} Larger blocks may take more time to propagate through the p2p network, leading to more forks~\cite{croman2016scaling, ren2019analysis}. However, in EIP1559, the block size is variable and dynamically adjusted; thus, its impact on the fork rate is not well understood. Our results empirically show that the London hard fork increased the block size on average and led to an approximately 3\% rise in fork rates. 
    \item {\em Network load.} We define the network load as the amount of computational, networking, and storage work that a node must perform to participate in the blockchain protocol. The community debated whether variable block sizes would increase the network load~\cite{eip_1559_spikes, buterin_2021_eip} since processing larger blocks consumes more resources. Our results show that EIP-1559 does not put the blockchain system under a significantly higher load for an extended period than the prior TPM . We do observe load spikes (periods during which an above-average amount of gas is consumed), but their frequency before or after the London fork is not significantly different.
    \item {\em Miner Extractable Value (MEV~\cite{daian_2019_flash}).} MEV refers to the profit that a miner can make through her ability to arbitrarily include, exclude, or reorder transactions within the blocks that she produces. Daian et al.~\cite{daian_2019_flash} point out that significant MEV can incentivize miners to deviate from the consensus protocol (e.g., to fork or even rewind the blockchain to collect profit in MEV~\cite{daian_2019_flash}), thus destabilizing consensus. Through our empirical analysis, we find that MEV becomes a much larger share of miner revenue under EIP-1559, mainly because the base fees are burnt. This might create an incentive for miners to invest more in MEV extraction.
\end{itemize}
\vspace{3mm}

The rest of the paper is organized as follows. \Cref{sec:related work} reviews the related works in three lines of literature. \Cref{sec:background} introduces the background and details of the EIP-1559 upgrade. \Cref{sec:data} introduces our data sources, which we use in \Cref{sec:results} to derive our empirical results. \Cref{sec:conclusion} discusses the results and concludes. Readers can refer to the working paper version on arXiv for an Appendix: https://arxiv.org/abs/2201.05574.

The datasets that we built in this paper might be of independent interest and have been released in~\cite{waitingtimedata} and~\cite{DVN/K7UYPI_2022}.

\section{Related Works}
\label{sec:related work}

This paper is related to three lines of literature: transaction fee mechanism design, waiting time modeling in market design,  and consensus security. 

\subsection{Transaction Fee Mechanism Design}

Since EIP-1559 was proposed, four recent papers have specifically investigated the proposal from different theoretical perspectives. Roughgarden (2021)~\cite{roughgarden_2020_transaction} provides a general framework for transaction fee mechanism design and proves that the EIP-1559 mechanism has incentive compatibility for myopic miners and off-chain agreement proofness. That is, myopic miners have incentives to act along with the allocation rules, and no off-chain agreement or collusion can give a higher return for miners. In addition to these results, Roughgarden (2020)~\cite{roughgarden_2020_transaction} analyzes the transaction fee and waiting time characteristics of EIP-1559 and points out that while no transaction fee mechanism can substantially lower transaction fees, EIP-1559 should lower the variance in transaction fees and waiting time through the flexibility of variable size blocks. The paper also argues that EIP-1559 does not weaken  system security regarding several types of attacks. Leonardos et al.~\cite{leonardos_2021_dynamical} put the EIP-1559 mechanism in a dynamic system framework and study the stability of the system. They show that the base-fee adjustment parameter is critical to system stability and provide threshold bounds for the adjustment parameter. Reijsbergen et al.~\cite{reijsbergen_2021_transaction} find that since the London hard fork, block sizes have intense and chaotic oscillations, which they believe could lead to harder fee estimation, and propose an additive increase and multiplicative decrease (AIMD) fee-adjusting model that can mitigate the spikes of block gas used.

Our work contributes to a growing economics and computer science literature on blockchain transaction fee mechanism design. The white paper of Bitcoin~\cite{nakamoto_2008_bitcoin} proposed the first-price auction mechanism for the Bitcoin Payment System (BPS), which was later widely adopted by other early blockchains (e.g., Ethereum before the London hard fork, Litecoin). Several papers analyze the supply and demand equilibrium of the BPS fee dynamics~\cite{houy_2014_the, rizun_2015_a, tsang_2021_the, ilk_2021_stability, pagnotta_2021_decentralizing}, while others analyze the game-theoretical equilibrium~\cite{dimitri_2021_view}. Alternative mechanisms for transaction fees have also been proposed. For example, Lavi et al.~\cite{lavi_2019_redesigning} and Yao~\cite{yao_2018_an} propose a monopolistic price mechanism where all transactions in the same block pay the same transaction fee, determined by the smallest bid. This approach is akin to the second-price auction. Basu et al.~\cite{basu_2021_stablefees} propose StableFees, a mechanism also based on a second-price auction with a more realistic model of miner behavior. Ferreira et al.~\cite{ferreira_2021_dynamic} propose a modification to EIP-1559 based on a dynamic posted-price mechanism that achieves more stability than EIP-1559 by their analysis. Li~\cite{li_2017_obviously} proposes a general concept of obviously strategy-proof (OSP) mechanisms that gives a rationale for providing more obvious mechanisms. Zhang and Levin~\cite{zhang_2017_bounded} further provide a decision theory foundation for the OSP mechanism for boundedly rational players.

\subsection{Waiting Time Modeling in Market Design}
Long waiting times and high transaction costs are major issues caused by network congestion, which is directly related to the scalability of blockchain~\cite{croman_2016_on, gudgeon_2020_sok}. Easley, O'Hara, and Basu~\cite{easley_2019_from} provide a game-theoretical model of the BPS with an important complication on mempool queuing that relates user welfare to fee levels and waiting time. Huberman, Leshno, and Moallemi~\cite{huberman_2021_monopoly} further link the BPS to monopoly pricing of miners and suggest a protocol design of adjustable system parameters for efficient congestion pricing, which coincides with the idea of EIP-1559. Waiting time auctions and market designs to minimize frictions have been extensively studied in economics and operation research~\cite{holt_1982_waitingline, roth_2002_the, niyato_2020_auction}. It is crucial to shorten waiting times, according to research in consumer psychology~\cite{leclerc_1995_waiting, kumar_1997_the} and the transient nature of many DeFi trading opportunties on blockchain~\cite{harvey_2021_defi}.

While the waiting time (delay) is widely used in theoretical models of users’ utility function, few have found an effective way to directly measure and analyze it in the blockchain setting. Some use an external data source on waiting time and mempool size that is only available for Bitcoin~\cite{easley_2019_from}; others use block size or fee levels as proxies for network congestion~\cite{huberman_2021_monopoly, sokolov_2021_ransomware}. Azevedo Sousa et al.~\cite{azevedosousa_2020_an} use an approach similar to our paper’s by directly observing the mempool of Ethereum, but their data suffer from the negative waiting time problem because of network latency. Our paper solves this problem by using the timestamp of the next block after the transaction concerned is included.

\subsection{Consensus Security}

The security of blockchain systems has been widely discussed since their inception~\cite{lin_2017_a}. Several papers analyze the incentive system of the Bitcoin system and propose potential attacks given specific incentive incompatibilities~\cite{eyal_2014_majority, lewenberg_2015_bitcoin, carlsten_2016_on, sompolinsky_2017_bitcoins, pass_2017_fruitchains, lehar_2020_miner}. Other studies extend the analysis to proof-of-stake protocols~\cite{chitra_2020_competitive, neuder_2020_selfish}.

The frequency of uncle blocks is an important indicator of blockchain forks that which endanger network security. Uncle blocks in the Ethereum community refer to blocks submitted for a block height after that block height is finalized and miners have moved to the next block height. Ethereum adopts a variation of the Greedy Heaviest Observed Subtree (GHOST) design~\cite{buterin_2013_ethereum, sompolinsky_2015_secure} that also provides block rewards to the miners of uncle blocks. Previous studies on Bitcoin show that a larger block size leads to a longer propagation time, making it more likely for some miners to submit an uncle block~\cite{decker_2013_information, sompolinsky_2015_secure}. A higher uncle rate can lead to less network resilience to double-spend attacks and selfish mining, thus endangering consensus~\cite{gervais_2016_on}.

Daian et al. ~\cite{daian_2019_flash} first introduced the potential impact of MEV on security. Many works have analyzed MEV extraction in various blockchain infrastructures ~\cite{angeris_2019_an, ante_2021_smart, bartoletti_2021_sok}. Chen et al.~\cite{chen_2020_a} investigate and systematize the vulnerabilities, attacks, and defenses of the Ethereum system security. Qin, Zhou, and Gervais~\cite{qin_2021_quantifying} quantify the specific value of MEV and provide evidence that mining pools are extracting MEV themselves. In early 2021, the inception of Flashbots made it easier to extract MEV and observe MEV extraction. In just a few months, the adoption rate of Flashbots increased rapidly and, at the time that this paper was written, was above 95\%~\cite{flashbots_2021_flashbotspm}. More MEV extraction tools and protocols have appeared recently, including the Eden network and Taichi network~\cite{Taichiprivatechannel, a2021_eden}.

\section{Background}
\label{sec:background}

\subsection{Transaction Fees in Ethereum}
It takes bandwidth, computational, and memory resources to successfully execute operations on the Ethereum network~\cite{buterin_2018_blockchain}. The amount of resource consumed is measured in the unit of gas. For example, it costs 21,000 gas to send a transaction and 53,000 gas to create a smart contract.\footnote{See Appendix G of Ethereum Yellow Paper,
\par
https://ethereum.github.io/yellowpaper/paper.pdf. 
}
To prevent malicious users from spamming the network or deploying hostile infinite loops, every operation is charged a fee~\cite{buterin_2013_ethereum}. The gas fee is paid in Ether,\footnote{The gas price is usually measured in Gwei and 1 Gwei $=  10^{-9}$ ETH.} dubbed ETH, the native currency of the Ethereum network, and calculated as:
\[
    \text{GasFee} = \text{GasUsed} \times \text{GasPrice}
\]
Sending a transaction could trigger a series of other operations on the Turing-complete blockchain. Therefore, the amount of gas needed for a transaction is usually unknown before execution. To avoid undue gas consumption, users can specify a gas limit with their transactions. Unconsumed gas is refunded. Ethereum had a block gas limit of 15 before the implementation of EIP-1559 that increased to 30 million after it. The sum of the gas limit of transactions included in a block cannot exceed this block gas limit. In the following two subsections, we explain the emergence of GasPrice    in the above equation before and after the London hard fork.

\subsection{Pre-EIP-1559 Transaction Fee Mechanism}
The pre-EIP-1559 legacy transaction fee mechanism is essentially a first-price auction. Users submit a gas price bid for their transactions to outbid competitors. Miners are incentivized to include those with the highest gas prices in a block first. However, the first-price auction does not have a dominant strategy equilibrium~\cite{mcafee_1987_auctions}, so users need to make assumptions about their competitors’ bids to optimize their bid strategy, a process that is impractical and user-unfriendly. In addition, bidding leads to distorted resource allocation, such as overpaid and volatile gas fees and unduly long inclusion times for transactions, which we examine later. To resolve these issues, a new gas fee mechanism was proposed, discussed, and implemented as EIP-1559.

\subsection{The new TFM in EIP-1559}
\label{subsec:background:eip1559}

EIP-1559~\cite{ethereumfoundation_2021_london} introduces four major changes to the transaction fee mechanism on Ethereum. A list of notations related to EIP-1559 is presented in~\cref{notation}.

\boldhead{Block Size}
EIP-1559 changes the fixed-sized blocks to variable-sized blocks. The block gas limit is doubled from 15 million to 30 million, while the block gas target is still set at 15 million. As we introduce below, a novel gas price mechanism ensures that the block gas used remains around the block gas target on average.

\boldhead{Base Fee} EIP-1559 introduces a base fee parameter determined by network conditions. The base fee is the minimum gas price that every transaction must pay to be included in a block. The base fee adjusts in a dynamic Markov process according to the block gas used in the previous block. If the block gas is greater than the target, the base fee for the next block increases, and vice versa. The base fee of the next block is determined solely by its present state. The dynamics of the base fee are represented as follows: 
\begin{equation}
      \text{BaseFee}_{h+1} = \text{BaseFee}_h (1+ \frac{1}{8} \frac{\text{GasUsed}_h - \text{GasTarget}}{\text{GasTarget}}).
      \label{base_fee_adjustment}
\end{equation}

Here, $h$ refers to the block height. BaseFee$_h$ and GasUsed$_h$ refer to the base fee and the block gas used in block $h$. GasTarget is fixed at 15 million.

\boldhead{User Bidding} How users bid is modified in a backward-compatible manner. Users can optionally bid two parameters in their transactions, the max priority fee per gas and the max fee per gas. Priority fees per gas are the tips with which users incentivize miners to prioritize their transactions. Max fees are the fee caps that users will pay including both base fees and priority fees. The difference between the max fee and the sum of the base fee and priority fee, if any, will be refunded to the user. The actual GasPrices of these transactions are calculated by:
$$\text{GasPrice} = \min \{ \text{BaseFee} + \text{MaxPriorityFee}, \text{MaxFee}\}.$$

For example, if a user bids $(\text{MaxFee}, \text{MaxPriorityFee}) = (60, 2)$, then there can be several cases depending on the level of base fee in the current block:
\begin{enumerate}
    \item If $\text{BaseFee} > 60$, the transaction must not be included in this block. It waits in the mempool until the base fee falls.
    \item If $58 < \text{BaseFee} < 60$, the miner can choose whether to include this transaction. If the transaction is included, then aside from the base fee, the user pays $60  \text{Basefee}$ Gwei as a priority fee to miners. Users pay $60$ Gwei per gas in total.
    \item If $\text{BaseFee} < 58$, the miner can choose whether to include this transaction. If the transaction is included, aside from the base fee, the user pays $2$ Gwei as a priority fee to miners. Users pay $\text{BaseFee} + 2$ Gwei per gas in total.
\end{enumerate}

It is worth noting that users are allowed to follow the legacy bid style and only bid a gas price, in which case the difference between gas price and base fee are all taken by miners as tips.

\boldhead{Miners' Revenue} The base fee is burned, while the priority fee is remitted to the miners as a reward. Before EIP-1559, miners earned all gas fees in a block. With EIP-1559 implemented, tips are de facto mandatory because miners do not earn the base fee; otherwise, they may mine empty blocks. Miner revenues include mainly static rewards,\footnote{The static reward is 2 Ether per block since the Constantinople fork in February 2019~\cite{ethereumfoundation_2019_constantinople}.} priority fees in the block, and uncle rewards if they mine an uncle block. In addition, miners receive profits extracted from including, omitting, ordering, and inserting transactions, known as the Miners Extractable Value (MEV)~\cite{daian_2019_flash}.

\begin{table}[htbp]
\setstretch{0.8}
    \centering
    \small
\begin{tabular}{c c} 
\textbf{Notation} & \multicolumn{1}{p{0.7\columnwidth}}{\textbf{Description}} \\ [0.5ex] 
\toprule
BaseFee & \multicolumn{1}{m{0.7\columnwidth}}{The minimum GasUsed multiplier required for a transaction to be included in a block. The result of BaseFee times GasUsed is the part of the transaction fee that is burned}\\
\midrule
MaxPriorityFee & \multicolumn{1}{m{0.7\columnwidth}}{The maximum GasUsed multiplier that a user is willing to pay to the miner}\\
\midrule
MaxFee & \multicolumn{1}{m{0.7\columnwidth}}{The maximum GasUsed multiplier that a user is willing to pay for a transaction}\\
\midrule
GasPrice & \multicolumn{1}{m{0.7\columnwidth}}{Only legacy transactions use it, which represents the GasUsed multiplier that a user is willing to pay for a transaction}\\
\midrule
GasUsed & \multicolumn{1}{m{0.7\columnwidth}}{The total amount of gas used by a transaction}\\
\midrule
GasTarget & \multicolumn{1}{m{0.7\columnwidth}}{The target of gas that blocks are expected to use on average, which is set by the protocol} \\
\midrule
GasFee & \multicolumn{1}{m{0.7\columnwidth}}{The actual transaction fee that a user pays} \\
\bottomrule
\end{tabular}
\caption{Notations related to EIP-1559} 
\label{notation}
\end{table}

\section{Data}
\label{sec:data}

\boldhead{Data availability} The final data records are stored and published on the Harvard Dataverse~\cite{DVN/K7UYPI_2022}.
\subsection{Data Sources and Metadata}
We use four data sources. First, we query the blockchain data from Google Bigquery, which documents the block-level characteristics and transaction-level characteristics from Ethereum~\cite{googlecloudplatform_2019_big}. Second, we run four Ethereum full nodes geographically distributed around the world (North Carolina, Los Angeles, Montreal, and Helsinki) to monitor the mempool of Ethereum constantly so that we can capture a historical log of the Ethereum mempool. Most users submit their transactions to the mempool so that miners can consider their transactions.~\footnote{There are transactions (e.g., Flashbot bundles) that bypass the mempool, but they are relatively rare as of the time of writing. According to the Flashbots API~\cite{flashbotsAPI},  around the time of the London hard fork, there were an average of 2.9 transactions per block from Flashbots bundles. Etherscan records private transactions in its perspective, and their number is also small.} The data fully capture the submission of each awaiting transaction in the mempool, including the time submitted and the bids on gas prices. It is worth mentioning that the Ethereum mempool data are ephemeral, so our data are not reproducible at a later time. Third, we query ETH price data at one-minute granularity from Bloomberg Terminal~\cite{bloomberglp_2019_xetusd}. We use these data to compute the minute-level price volatility of ETH prices as a control variable, which is an instrument for the demand for transactions on Ethereum. Fourth, we use the Flashbot API to collect the miner revenues, including Flashbot revenues, in each block. A data dictionary can be found in Appendix~\cref{Appendix A}.

\begin{figure}
      \centering
      \includegraphics[width = \linewidth]{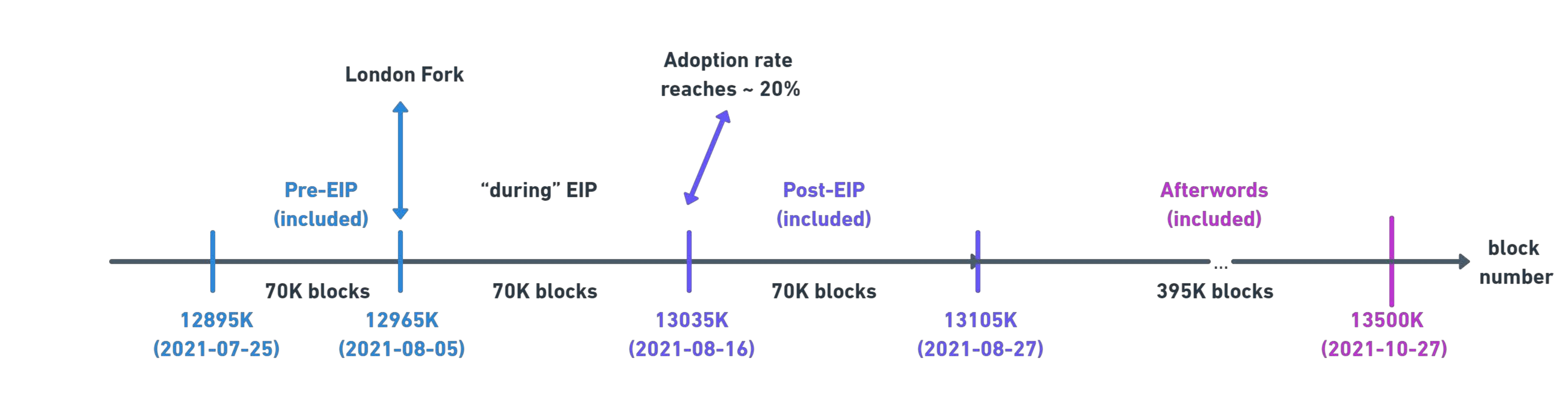}
      \caption{Periods and Block Numbers}
      \label{fig:timeline}
  \end{figure}

The time of our data is specified in~\cref{fig:timeline}. For the pre–London hard fork period, we use data from block numbers 12895000 (2021-07-20) to 12965000 (2021-08-05, the block of the London hard fork). For the post-EIP-1559 period, we use data from block number 13035000 (2021-08-16) to block number 13105000 (2021-08-31). We do not use data from the blocks immediately after EIP-1559 because it took time for users and miners to upgrade their software to adapt to the London hard fork change. We set the Start Block of the post-EIP 1559 period to the block at which adoption reaches 20\% . We do not record mempool data from between 2021-08-05 and 2021-08-16.

\subsection{Transaction Fee Data}\label{subsec:txfee}
We obtain blockchain data including the gas price paid for each transaction (legacy and EIP-1559) and the max fee and max priority fee bids for each EIP-1559 transaction from Google Bigquery~\cite{googlecloudplatform_2019_big}.

From transaction-level data, we derive several block-level metrics. To measure the gas price of a ``representative transaction'' in the block, we use the median of gas prices from all transactions.
Measuring the "variation" in gas prices in the block is somewhat trickier since there are outliers. Standard measures such as the mean and standard deviation can be misleading and statistically meaningless. However, measures such as the median and interquartile range (IQR) are much more robust to outliers, which we encounter frequently in blockchain data. We use the standardized IQR of gas prices in blocks to measure the intrablock variance of gas prices, defined as
\[
\text{standardized IQR} = \frac{Q_{75}(\text{GasPrice}) - Q_{25}(\text{GasPrice})}{Q_{50}(\text{GasPrice})}
\]
where $Q_{25}$, $Q_{50}$, and $Q_{75}$ refer to the 25th, 50th, and 75th percentiles of the gas prices paid in a specific block. 
In addition, our data include block-level blockchain data such as the miner's block timestamp, base fee per gas, block gas used, and block size.

\subsection{Waiting Time Data}
\label{subsec:wtdata}

\newcommand{\tx}{\ensuremath{\mathsf{TX}}}
\newcommand{\txblockchaintime}{T^\tx_\text{block}}
\newcommand{\txmempooltime}{T^\tx_\text{mempool}}

We define the waiting time of a given transaction $\tx$ as the time that the transaction waits in the mempool, namely $$\text{Waiting time of \tx}=\txblockchaintime-\txmempooltime$$
where $T^\tx$  is the time when the transaction {\em first} appears in the mempool and $\txblockchaintime$ is the time when the transaction is mined.

\boldhead{Estimating $T^\tx$}
Obtaining an exact $T^\tx$  is challenging because that would require monitoring the traffic of all (or most) P2P nodes.
To obtain a reasonable approximation, we place probing nodes across the globe to obtain representative samples of the mempool. Specifically, we modify the Ethereum Geth client~\cite{geth} so that our nodes connect to up to 1,000 peers. We deploy $4$ geographically distributed nodes in Durham, Los Angeles, Montreal, and Helsinki. 
Our modified Geth client stores a log of mempools whenever they receive a new transaction from the P2P network. We use the earliest time when $\tx$ is observed in the mempool across all servers as the estimate of $T^\tx$.Although we make various efforts to improve the accuracy of the mempool observations by placing probing nodes in different locations, increasing their connectivity, and connecting to well-known nodes such as established mining pools ~\cite{deter},
we note that the estimation may not be  perfect. Further improving the mempool observations could be an interesting subject of future work.

\boldhead{Estimating $T^\tx$}
As defined above, $T^\tx$ is the time when the transaction appeared in a block. One may attempt to set $T^\tx$ to the block timestamp given by miners, but that is vastly inaccurate. For instance, Azevedo-Sousa et al.~\cite{azevedosousa_2020_an} calculate the waiting time in this fashion, leading to the wrong conclusion that 50\% of transactions have {\em negative} waiting times!
The reason is that the block timestamps given by miners are typically when the miner {\em starts} the mining process whereas $T^tx$ is when the mining process {\em ends}.

Obtaining a precise $T^\tx$ would require monitoring the traffic 
of all (or most) miners, which is very challenging. We bypass this difficulty by using the timestamp of the  {\em next} block (i.e., the block after the one in which $\tx$ appeared) as an approximation because the next miner usually begins the mining process as soon as they receive the previous block to maximize the success rate. 

By the above steps, we reduce the proportion of transactions with negative waiting times from 50\% (as encountered in~\cite{azevedosousa_2020_an}) to less than 1\%: the percentage of negative waiting time in our data is 0.4\% during "Pre-EIP," 0.3\% during "Post-EIP," and 0.8\% during "Afterwords," respectively, for the periods shown in Figure~\ref{fig:timeline}. The remaining negative waiting time may be caused by inaccurate block timestamps (some miners may add a wrong timestamp accidentally or maliciously, although it cannot deviate from real time too much or the block will be ignored by honest nodes~\cite{geth_future_time}) or errors in estimating $T^\tx$ (see above). Therefore, we set the waiting time of those transactions to 0. 
We aggregate the waiting time to the block level by taking quartiles in a manner similar to how we aggregate gas prices. The median waiting times that we mention below are block-level statistics, representing the median of all transaction waiting times in a specific block.

\subsection{Miners' Revenue Data}
\label{subsec:revenue data}
To investigate how EIP-1559 changes miners’ incentives, we collected data about miner revenue.

Miner Revenue (MR) includes block rewards, transaction fees, and "extracted values" (MEV~\cite{daian_2019_flash}). We can observe the first two components of MR on the blockchain, but it is difficult to completely capture MEV (though previous works, e.g.,~\cite{qin_2021_quantifying}, looked into MEV from specific attacks). We use the revenue from Flashbots, by far the largest MEV extraction services~\cite{flashbots_2021_flashbotspm}, as an approximation of the total MEV.

Flashbot revenue comes in two forms: gas fees of transactions in so-called Flashbot Bundles (FBBs) and direct payments to the miners by transactions in FBBs (it is typical for FBB transactions to pay miners by transferring ETH to the address of the miner who mines the block). Information about FBBs is publicly available through the Flashbots API~\cite{flashbotsAPI}.

To observe the long-term effect on Miner Revenue (MR), we collect data in a longer window than that in~\cref{fig:timeline}, from block numbers 12,710,000 to 13,510,000 (800,000 blocks in total), i.e., from 40 days before the London hard fork to 95 days after. Specifically, we divide miner revenue into five categories:

\begin{enumerate}[leftmargin=*]
    \item Static block rewards: 2 ETH per block
    \item Uncle inclusion rewards: $\frac{1}{32}$ ETH for referencing an uncle block
    \item Non-FBB gas fees: total gas fees of transactions not in FBBs
    \item FBB gas fees: total gas fees of transactions in FBBs
    \item FBB coinbase transfer: total amount of direct payments in FBBs
\end{enumerate}

As noted above, this division scheme means that we use the revenue from Flashbots as an approximation for MEV and use all the revenue that we can observe as an approximation for miner revenue.

\subsection{Fork Rate Data}
To understand how the new transaction fee mechanism may affect consensus security, we collected data about past forks in Ethereum. The Ethereum blockchain contains pointers to uncle blocks, from which we derive the number of ``siblings'' to show the specific time when forks happen. Specifically, a sibling of a block at height $h$ refers to the uncle blocks (of a later block) with height $h$. The sibling count can reflect how many different blocks compete at a specific height at a given time.

\subsection{Preliminary Visualizations}

We first visualize some parameters related to EIP-1559 here.

\begin{figure}
      \centering
      \subfigure[The base fee oscillated after the London hard fork with occasional peaks. Each dot represents a block.]{\includegraphics[width = \linewidth]{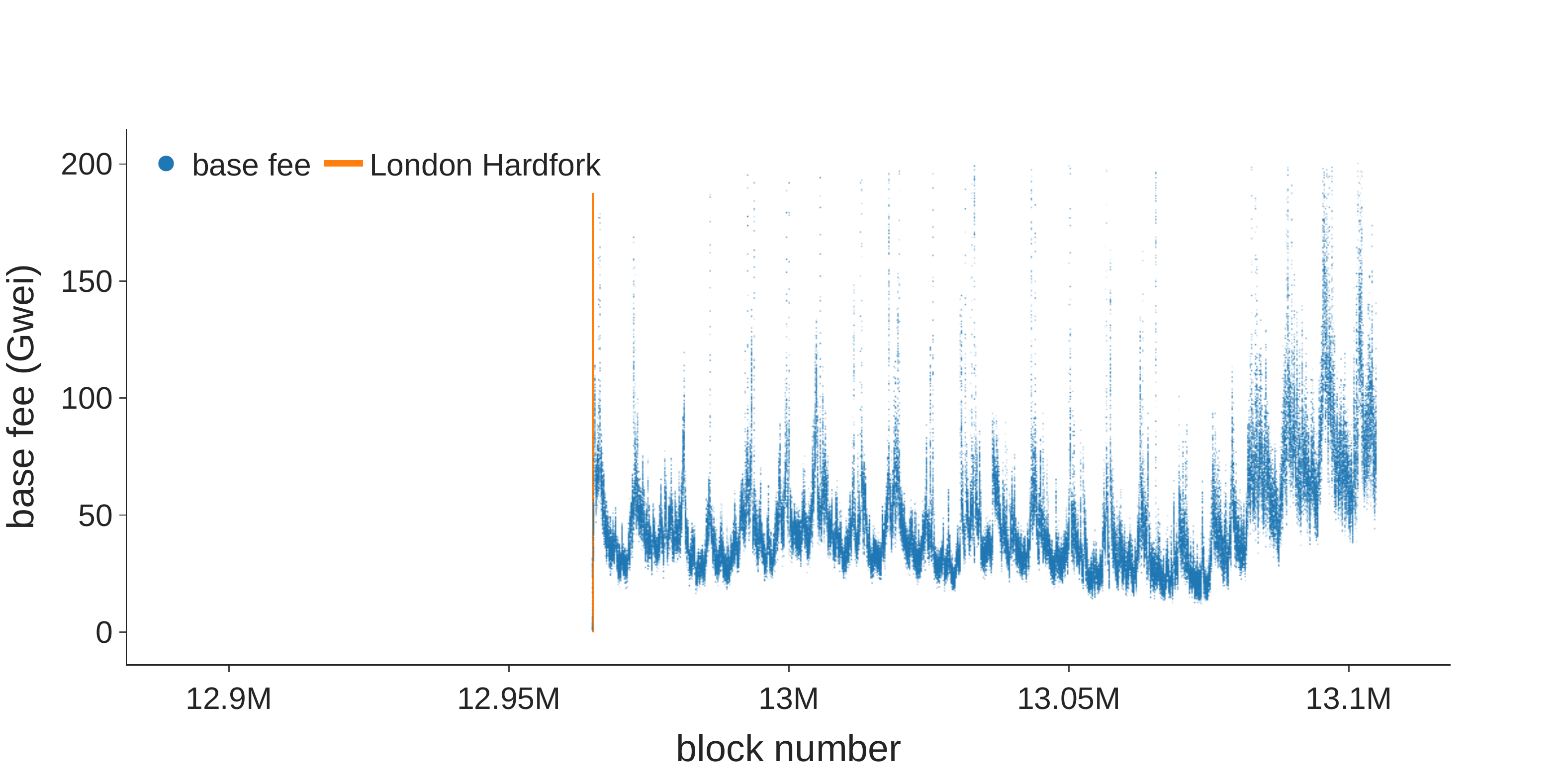}\label{fig:base fee}}
      \vspace{-12pt}
      \subfigure[The net supply of ETH dropped after the London hard fork, sometimes to negative levels. Netsupply is the number of ETH issued to miners minus that burned as base fees. Each dot represents a block.]{\includegraphics[width = \linewidth]{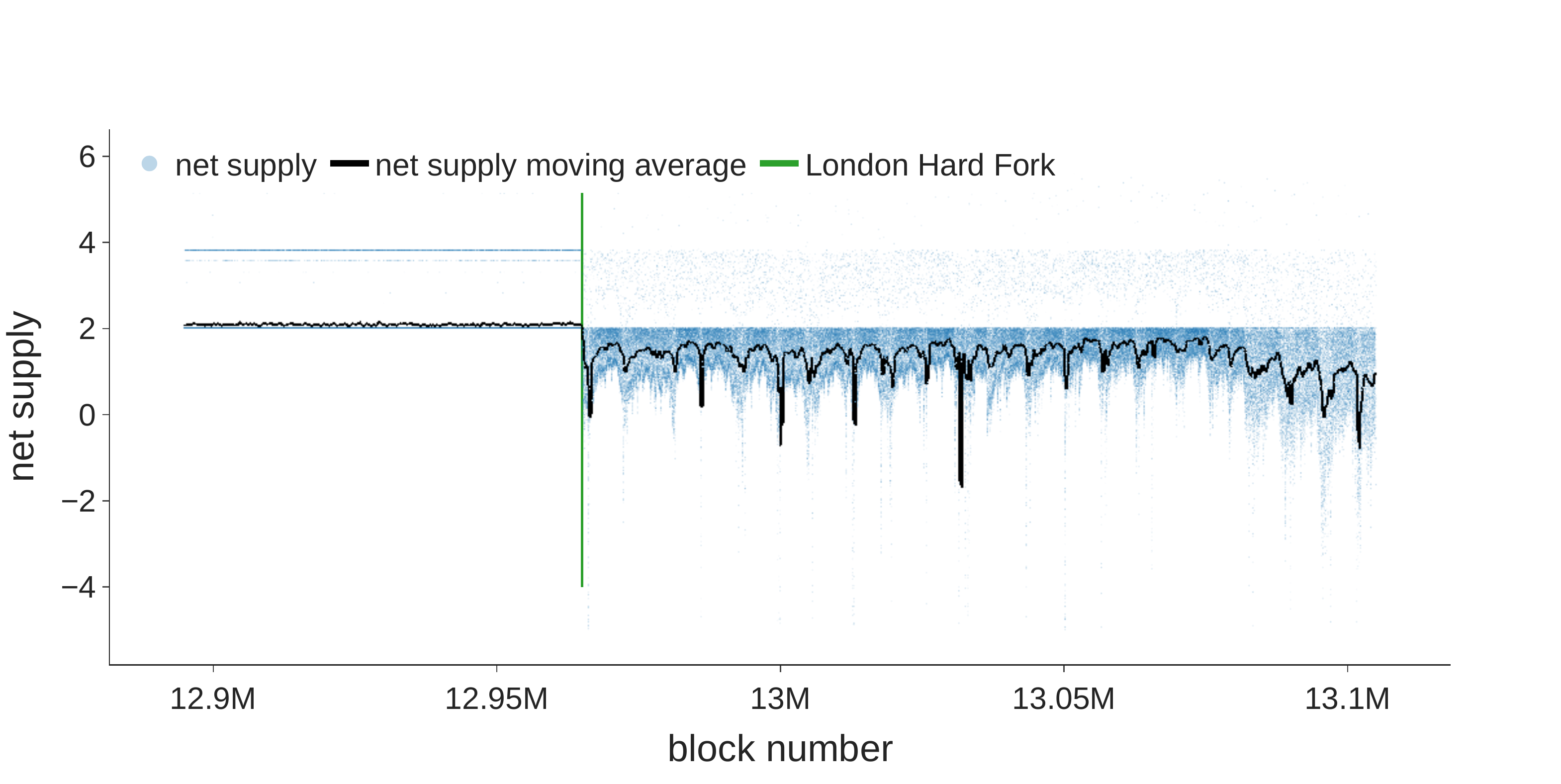}\label{fig:net supply}}
      \caption{Time Series of Base Fee and ETH Net Supply}
  \end{figure}

\boldhead{Base fee dynamics}
\Cref{fig:base fee} shows that the base fee oscillated between 30 and 200 Gwei after the London hard fork with occasional peaks in high-usage periods. With a significant amount of Ether burned as a base fee, the issuance rate of Ether reduces significantly. In certain circumstances, blocks could have more Ether burned than minted, resulting in a negative supply of Ether, as shown in~\cref{fig:net supply}. Base fee burning can create positive feedback between Ethereum network activity and the Ether price. High demand for Ethereum resources from users drives up both block gas usage and the base fee, which burns more Ether. The reduction of Ether supply induces bullish market sentiment, Ether price appreciation, and ultimately more users. As a result, the reduction in revenue from transaction fees might be partly offset by a higher Ether price~\cite{kim_2021_an}.

\begin{figure}
    \centering
    \subfigure[Before the London hard fork, almost all blocks used 15 million gas; since the London hard fork, the block gas used varies between 0-30 million. Each dot represents a block.]{\includegraphics[width=\linewidth]{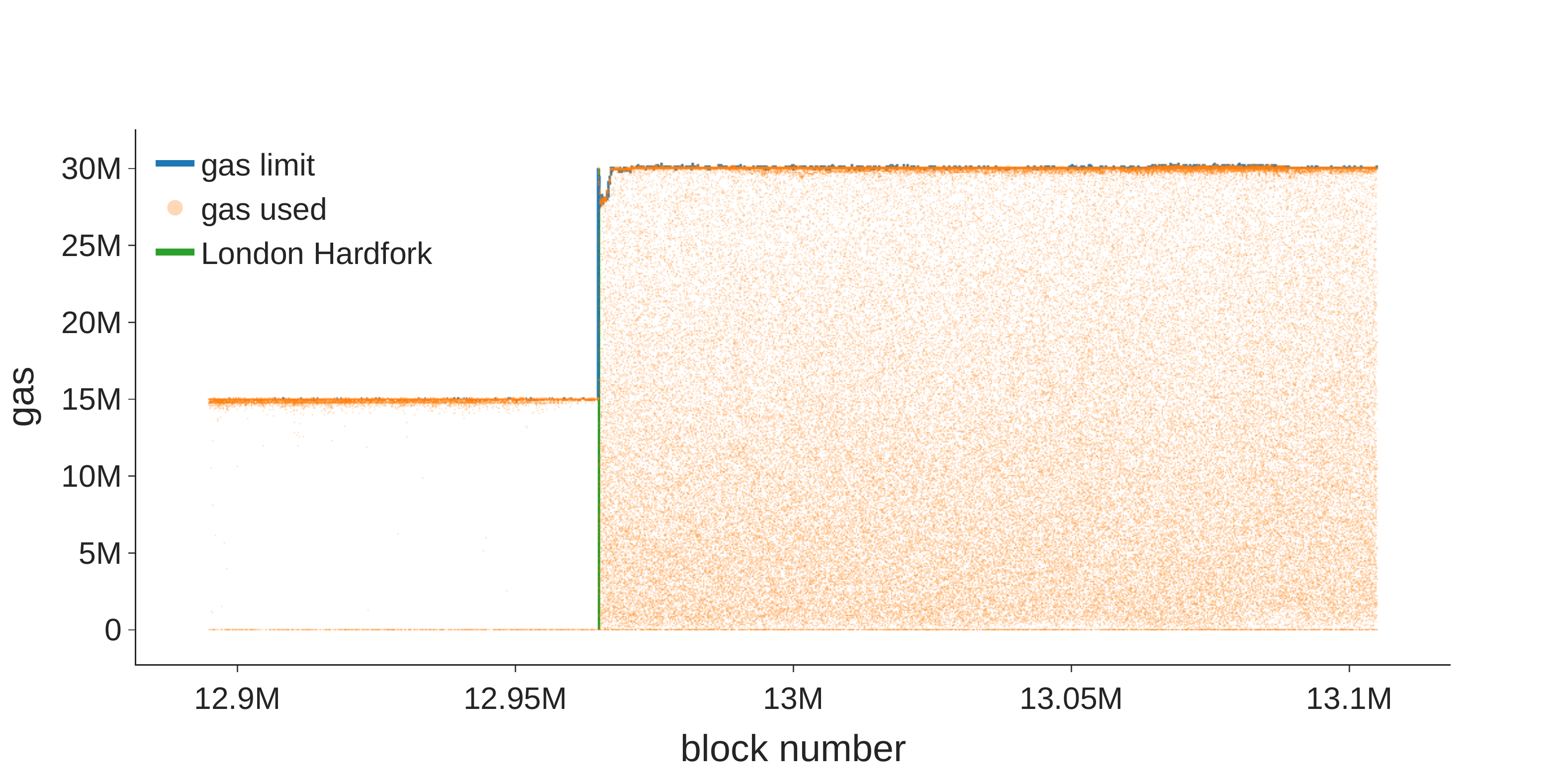}\label{fig:gasused1}}\\
    \vspace{-12pt}
    \subfigure[Approximately 20\% of blocks after the London hard fork are full (i.e., consume 30 million gas).]{\includegraphics[width=\linewidth]{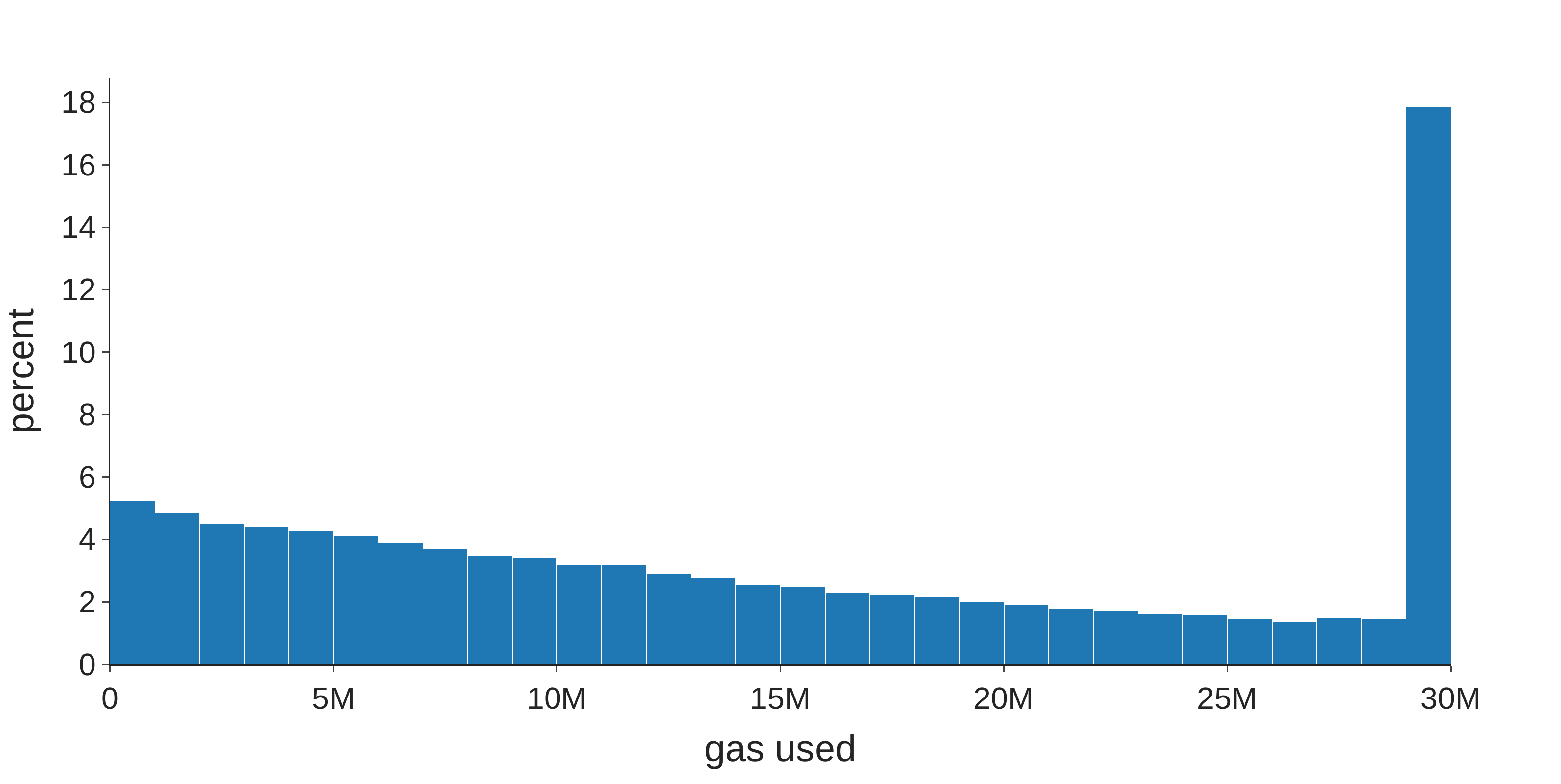}\label{fig:gasused2}}
    \caption{Distribution of Block Gas Used}
    \label{fig:gas used}
  \end{figure}

\boldhead{Block gas usage}
The new transaction fee mechanism led to increases and decreases in the base fee, as shown above, and volatile block gas usage, as shown in~\cref{fig:gas used}. In periods of high demand (e.g., Non-fungible token airdrops, market crashes~\cite{young_2021_nft, bestowie_2021_heres}), block gas used can deviate from its target of 15 million to at most 30 million (or slightly above), in which case the base fee in the next block will increase by at most 12.5\% of that of the current block as implied by~\cref{base_fee_adjustment}. Given the current block time of approximately 13 seconds, the base fee will double every 80 seconds if a series of full blocks are produced. The surge in the  base fee ensures that the limited block space is allocated to transactions with higher intrinsic values.
Increasing the base fee screens users with lower intrinsic values and leads to fewer transactions included in a block until the block gas used is lower than the target. 

\begin{figure}
      \centering
      \includegraphics[width = \linewidth]{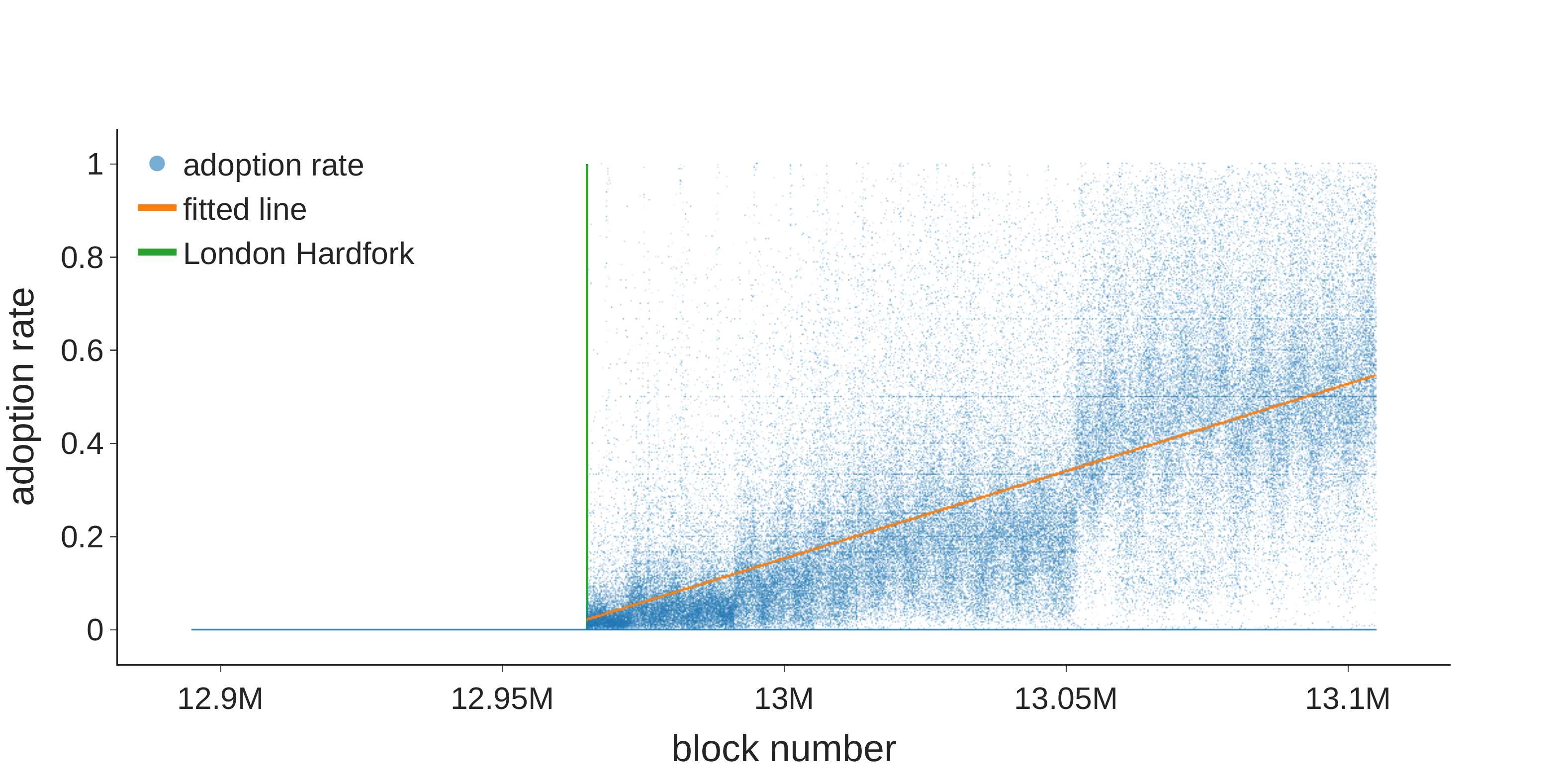}
      \caption{Adoption of EIP-1559-style bidding steadily increased after the London hard fork. Each dot represents a block.}
      \label{fig:adoption rate}
  \end{figure}

\boldhead{Adoption rate}
\Cref{fig:adoption rate} shows that the adoption rate of the new transaction fee mechanism has been steadily increasing. Transactions that adopt the EIP-1559 bidding style with the max fee and max priority fee are defined to be type-2 (TxnType = 2), while those that stick to the legacy bidding style are defined to be type-0 or type-1 (TxnType = 0 or 1), depending on which points on the elliptic curve are used. We notice a sharp increase in the adoption rate around block number 13.05 million, which is possibly related to the adoption as default on MetaMask~\cite{a2021_metamask}.

\section{Empirical Results}
\label{sec:results}
In this section, we first present an overview of our methodology in~\cref{subsec:methodology}, and then we present empirical results to answer the following questions:
\begin{itemize}[leftmargin=*]
    \item Does EIP-1559 affect transaction fee dynamics in terms of the overall fee level, users' bidding strategies, and the intrablock distribution of fees? (\cref{subsec:fee})
    \item Does EIP-1559 affect the distribution of transaction waiting time? (\cref{subsec:waitingtime})
    \item Does EIP-1559 affect consensus security, in terms of fork rates, network loads, and MEV? (\cref{sec:security})
\end{itemize}

\subsection{Methodology}
\label{subsec:methodology}
\boldhead{Code Availability} The code used for the empirical analysis is available on GitHub: \url{https://github.com/SciEcon/EIP1559}.

A key challenge in identifying the effect of EIP-1559 on blockchain characteristics is that we must isolate the effect of EIP-1559 from the effects of confounding factors, such as price volatility, network instability, and the time trend. Our approach is to adopt an event study framework~\cite{mackinlay_1997_event} and regression discontinuity design (RDD)~\cite{athey_2017_the, lee_2010_regression} to identify the impact of EIP-1559 on the dynamics of Ethereum. RDD is a quasi-experimental evaluation method widely used in economics, political science, epidemiology, and related disciplines for causal inference on the impact of an event,\footnote{We refer the readers to pages 5-8 in Athey and Imbens~\cite{athey_2017_the} for more details on the regression discontinuity design.} here the implementation of EIP-1559. We exploit the gradual adoption of EIP-1559 several weeks after the London hard fork to set up the RDD framework, using the event of the London hard fork and EIP-1559 adoption rate in each block as the independent variables and estimating both the immediate effects of the London hard fork and the average treatment effects of EIP-1559 adoption.

We specify the RDD by~\cref{eqn:reg}:
\begin{equation}\label{eqn:reg}
Y = \alpha_0 + \alpha_1 \mathbbm{1}(\text{London hard fork}) + \alpha_2 r_{\text{EIP}} + \mathbf{\alpha_3 X} + \mu_h + \epsilon.
\end{equation}

Here, $\alpha_1$ is the coefficient for the indicator variable for the occurrence of the London hard fork (affecting block numbers $\geq$ 12965000). It represents the immediate effect of EIP-1559 on the outcome variable $Y$. $\alpha_2$ is the coefficient for  $r_{\text{EIP}}$, the percentage of transactions adopting EIP-1559 after the London hard fork. Since EIP-1559 is backward compatible, many users still adopted the legacy bidding style in the few weeks after the upgrade, but the adoption rate kept rising as we show in~\cref{fig:adoption rate}. By Nov. 2021, 40\% to 60\% of all transactions~\cite{kim_2021_an} used the new bid style. Therefore, $\alpha_2$ represents the effect of an increase in EIP-1559 adoption.

We include a set of control variables~\cite{blackwell_2018_how} represented as $\mathbf{X}$ in~\cref{eqn:reg}. We control for the block number in our sample to account for a possible time trend before the London hard fork, defined by %
\[
\text{nblock} =
\begin{cases}
 \text{BlockNumber} - 12895000 & \text{pre-EIP period}  \\ 
 \text{BlockNumber} - 12965000 & \text{post-EIP period}
\end{cases}
.
\]
We also control for price volatility, median gas price, and return on investment (ROI) (the minute-level percentage change in the ETH price) to account for the variance in demand for transactions on the Ethereum network. Block size is controlled as a proxy of network stability. We include an hour fixed-effect term $\mu_h$ to account for the seasonality of Ethereum network conditions as we can clearly see in~\cref{fig:base fee}. $\epsilon$ is an error term.

\subsection{Transaction Fee Dynamics}
\label{subsec:fee}
\subsubsection{Overview}
Transaction fee mechanism design is not intended to solve blockchain scalability. Thus, the fee level before and after the London hard fork did not change substantially. However, it did change how users bid and with users' bidding strategy largely coinciding with the predictions made by Roughgarden (2020)~\cite{roughgarden_2020_transaction}. 

\begin{figure}[htp] 
\centering
  \subfigure[The gas price paid by users did not change much immediately after the London hard fork, but it started to rise approximately two weeks later. We cannot conclude whether this was caused by EIP-1559. Each dot represents a quartile of the gas price paid for a block.]{\includegraphics[width=\linewidth]{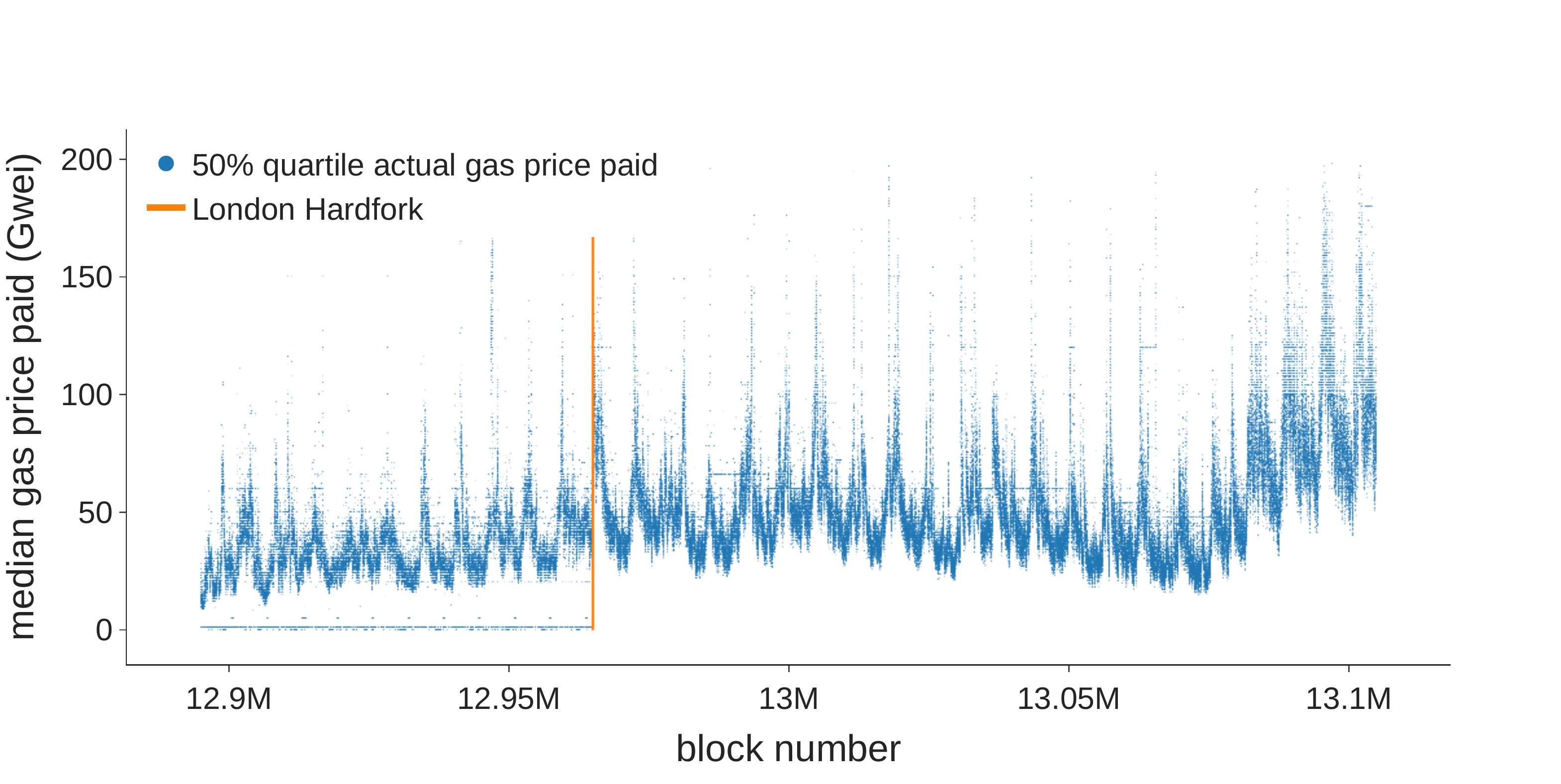}\label{fig:gas price}}%
  \vspace{-12pt}
  \subfigure[Max fee bids were usually much higher than the actual gas prices paid. Each dot represents a block.]{\includegraphics[width=\linewidth]{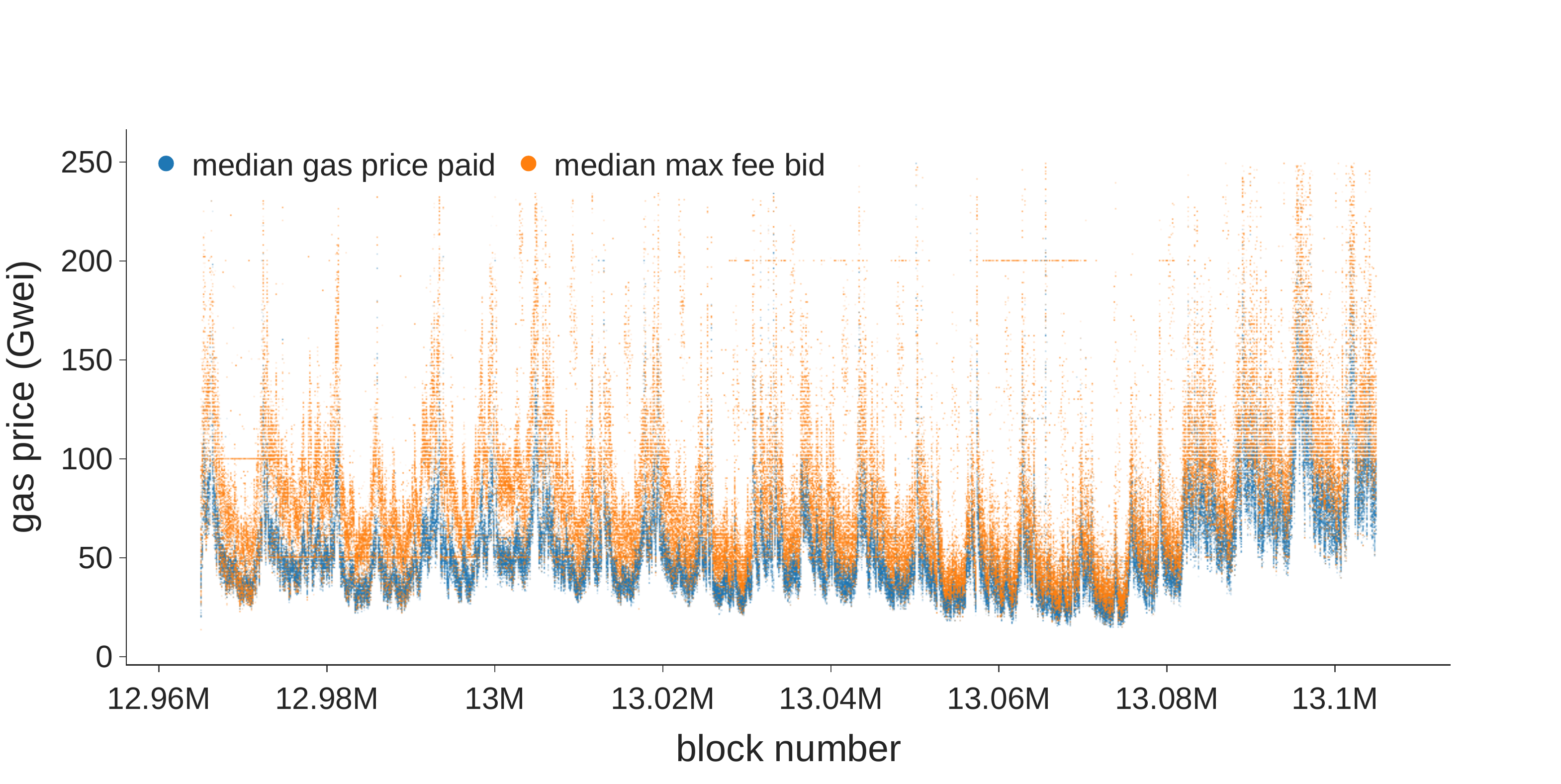}\label{fig:max fee & gp}}%
  \vspace{-12pt}
  \subfigure[Priority fee bids remained at a low level ($<$ 10 Gwei) in most cases, especially after block number 13.06 M. Each dot represents a block.]{\includegraphics[width=\linewidth]{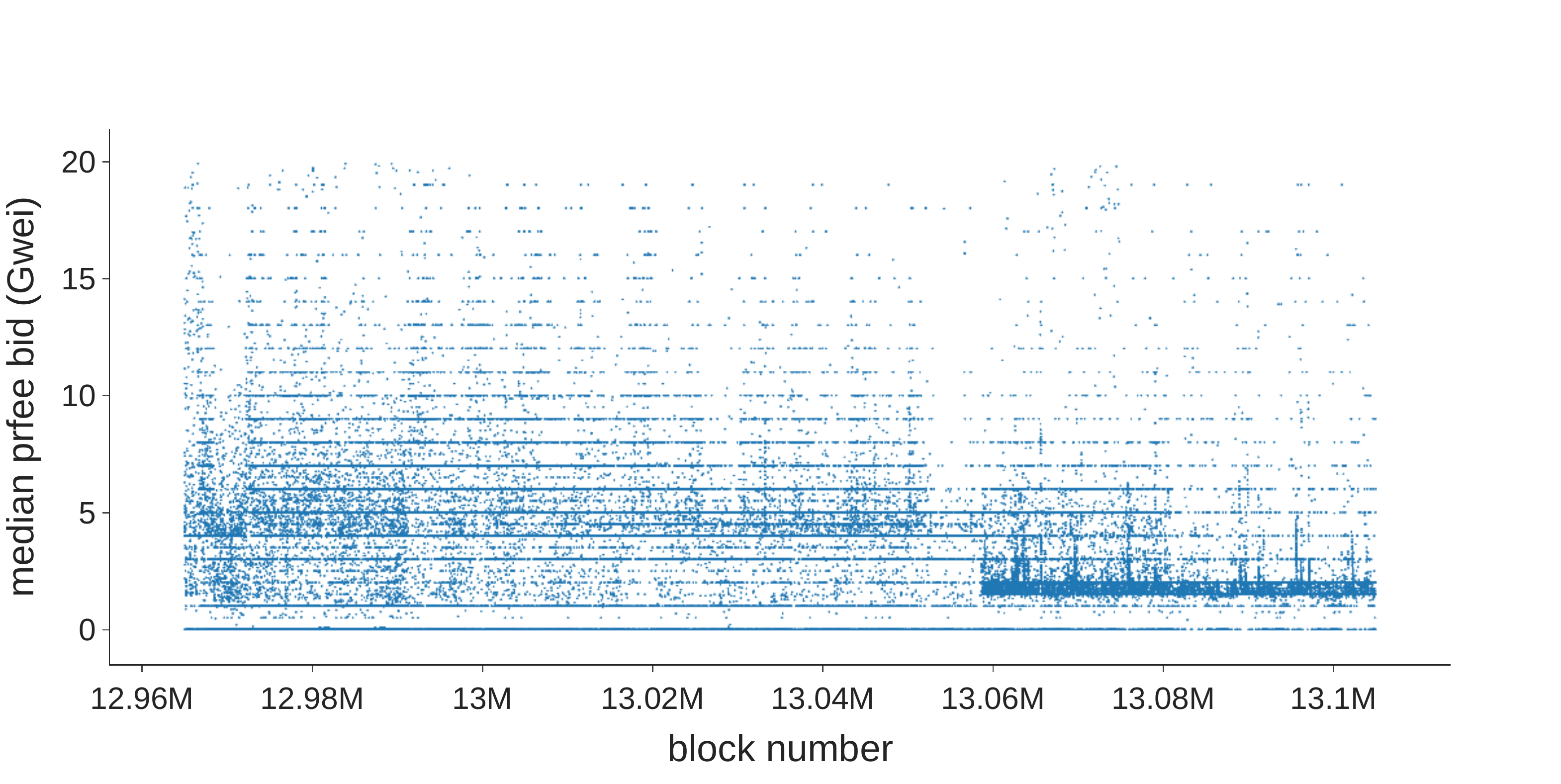}\label{fig:pr fee}}%
\caption{Overview of Fee Dynamics}
\end{figure}

\Cref{fig:gas price} shows the time series for the actual gas price paid quartiles in each block before and after the London hard fork. With hourly seasonality (intraday oscillation due to the difference in demand across time zones), the gas price level did not change much before or immediately after the London hard fork. It is unclear whether the gas price increase after block number 13.07M was caused by EIP-1559 or other factors.

\Cref{fig:max fee & gp} and~\cref{fig:pr fee} further decompose different fee parameters in users' bids. \Cref{fig:max fee & gp} shows that while the median gas price paid and median max fee bid are volatile and highly correlated with each other, the max fee bids are usually higher than the gas prices paid. Meanwhile,~\cref{fig:pr fee} shows that the median max priority bid remains at a low level (almost always $<10$ Gwei throughout the period and $<3$ Gwei after block number 13.06M). Overall, these results are consistent with the predictions in~\cite{roughgarden_2020_transaction} that the obvious optimal bid is a max fee that represents the intrinsic value of the transaction and a max priority fee that represents the marginal cost of miners' inclusion of the transaction. If these predictions hold, then we should observe max fee bids higher than the actual price paid and a low, stable level of priority fees bid, which is exactly what we observe. Ferreira et al.~\cite{ferreira_2021_dynamic} express concerns that the EIP-1559 TFM may degrade to a first-price auction on the priority fee when the base fee is set too low. However, our empirical results show that this did not happen much in practice.

\begin{figure}
      \centering
      \includegraphics[width = \linewidth]{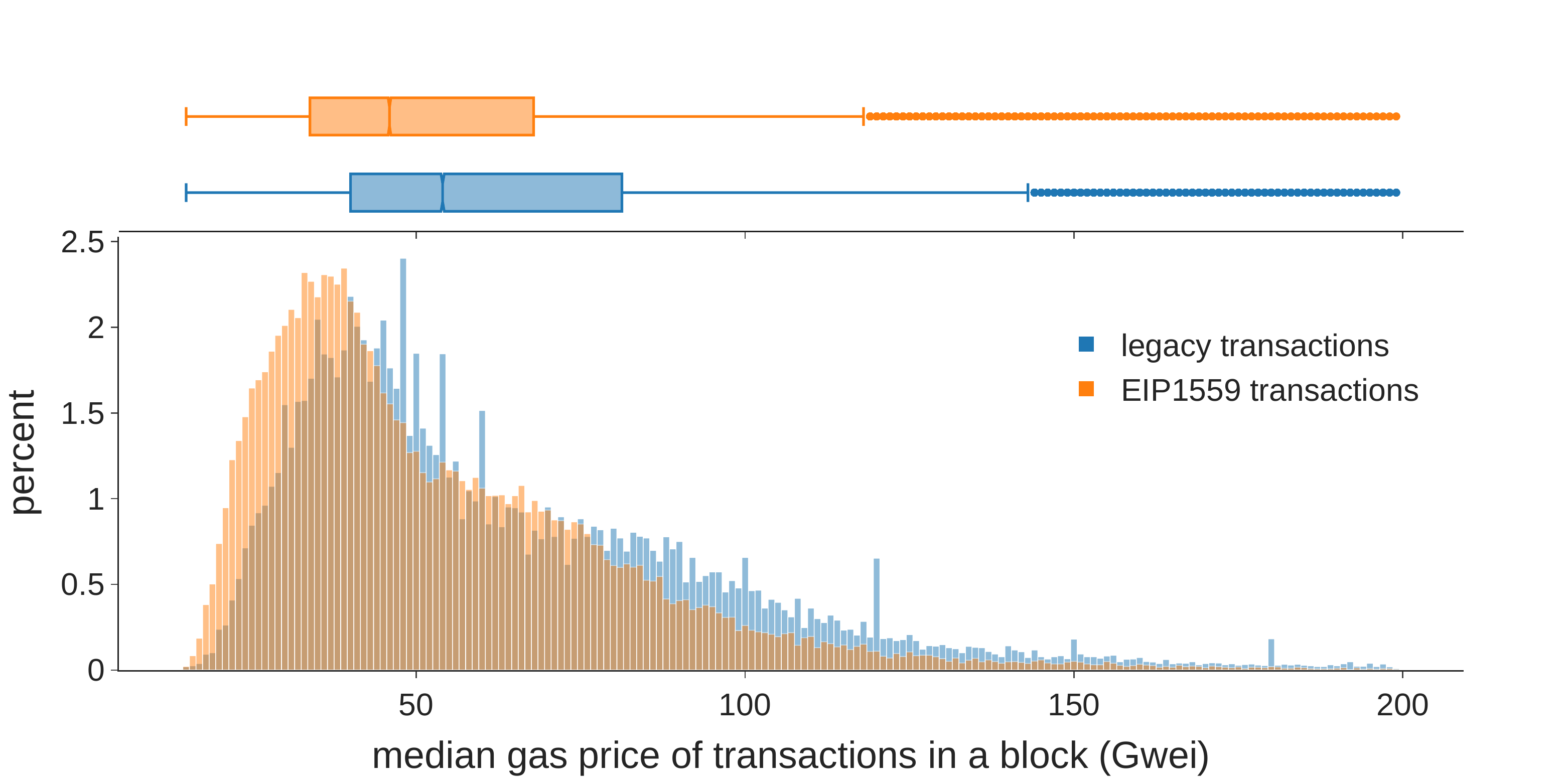}
      \caption{Distributions of median gas prices of a block for legacy transactions and EIP-1559 transactions. The distribution of EIP-1559 transactions is overall to the left of that of legacy transactions, which means that users who adopt EIP-1559 pay a lower gas price.}
      \label{fig:gpdistr}
  \end{figure}

Moreover, we compare the median prices for different transaction types. As shown in~\cref{fig:gpdistr}, the median gas price paid of the EIP-1559 transactions in a block has a distribution to the left of that of the legacy transactions. The median  gas price of EIP-1559 transactions in each block is 45 Gwei, while that of legacy transactions is 54 Gwei. This means that users who adopt EIP-1559 bidding overall pay less than those who stick to adhere to legacy bidding.

These findings point to the insight that users' fee estimation was made easier by EIP-1559. Before EIP-1559, users had to pay the entirety of the bid if their transactions were included, so they risked overpaying gas fees if the network was not as congested as they expected. Due to this concern, users had to carefully estimate gas prices and might have tended to bid conservatively as a result of the first-price auction~\cite{niyato_2020_auction}. EIP-1559, however, allows users to set two parameters in a bid: a priority fee and a max fee (both per gas unit). Users pay only the smaller of 1) the sum of the base fee and priority fee bid and 2) the max fee. Therefore, if the network conditions are better than expected, the user will pay only the base fee plus the priority fee. This leads to a simple but optimal bidding strategy, as shown in~\cite{roughgarden_2020_transaction}. Therefore, EIP-1559 makes fee estimation easier.
  
\subsubsection{Intrablock interquartile range}~\label{subsec:IQR}
If users bid in the obvious optimal bid introduced above, the eventual gas price paid by users who adopt the EIP-1559 style bid is the base fee plus a small priority fee, so users who adopt the EIP-1559 style bid tend to pay a similar gas price. A direct implication of this is that the intrablock variance of gas prices should have significantly declined after the London hard fork, especially as more users adopted type-2 transactions. While we cannot directly measure the ease of fee estimation, we manage to measure the intrablock variance of gas prices by the standardized interquartile range (IQR) (defined in~\cref{subsec:txfee}).

The relationship between the standardized IQR, the London hard fork, and EIP-1559 is visualized in~\cref{fig:cf_iqdiff}, which simulates the standardized IQR at different levels of EIP-1559 adoption rates based on Column (4) in~\cref{table: IQR}. For the period immediately after the London hard fork, the standardized IQR increases by approximately 8 percentage points (from 0.26 to 0.34), and the standardized IQR is predicted to decrease by 24 percentage points (from 0.34 to 0.10) when all users adopt the new bidding style.

Given the scale of this estimate, EIP-1559 should have a large negative effect on the intrablock difference of gas price paid as more users adopt the new bidding style (by Nov. 2021, the EIP-1559 adoption rate was approximately 40\%-60\%~\cite{kim_2021_an}). This implies that the inequality of intrablock gas prices decreases, especially as more users adopt EIP-1559 transactions. Thus, we show that the concerns raised by Reijsbergen et al.~\cite{reijsbergen_2021_transaction} relating to base fee volatility making fee estimation more difficult does not hold in practice.

\subsection{Waiting Time}
\label{subsec:waitingtime}

The waiting time is widely modeled in the literature as an essential component of users' utility function, and a short waiting time is crucial to the user experience. As a result of easier fee estimation (see above in~\cref{subsec:fee}), a rational user can simply bid her intrinsic value of the transaction without risking overpaying. Thus, with the EIP-1559 upgrade, it should be more straightforward for users to include their transactions in the next available block. We find that the waiting time significantly declined after the London hard fork. This benefits both EIP-1559 and legacy transactions.

\begin{figure}
  \centering
  \includegraphics[width = \linewidth]{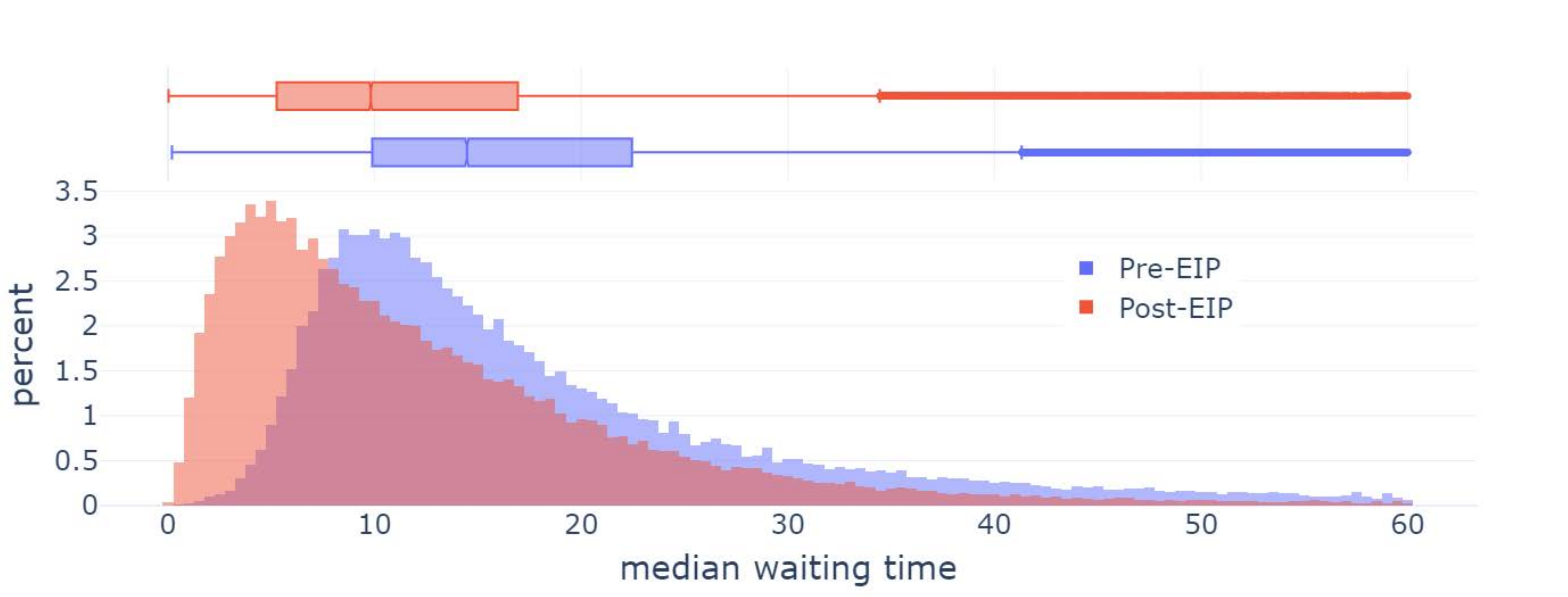}
  \caption{Distributions of median waiting time. The distribution moved left after the London hard fork. Users experience a much lower transaction waiting time with EIP-1559.}
  \label{fig:wtdistr}
\end{figure}

The reduction in the waiting time can be observed in~\cref{fig:wtdistr}, which demonstrates the distribution of the block median waiting time before and after the London hard fork. Each observation represents a block and the median of transaction waiting times in that block. The distribution shifted leftward after the London hard fork. The 25th quartile of the median block waiting time decreases from 10.7 seconds to 5.5 seconds, the 50th quartile decreases from 16.9 seconds to 10.4 seconds, and the 75th quartile decreases from 34.0 seconds to 18.6
seconds. Moreover, we observe that the waiting time of transactions of both types (legacy or EIP-1559 bidding styles) decreases. The 50th quartile of the median legacy-style transaction waiting time across blocks is 9.4 seconds for the period after the London hard fork, while that of the median EIP-1559-style transaction waiting time across blocks is 8.9 seconds.\footnote{Readers may notice that both median legacy-style transaction waiting time and median EIP-1559-style transactions waiting time are shorter than median all transaction waiting time after London hard fork. This does not contradict the fact that in each block, the median waiting time of all transactions is always between the median waiting time of legacy transactions and the median waiting time of EIP-1559 transactions. The distributions of these three variables are shown in~\cref{Appendix D} \cref{fig:wtdistrtype}}. This implies that the effect of EIP-1559 adoption spills over even to transactions not adopting the new bidding style by improving the overall gas bid structure in the mempool.
\Cref{table: waiting time} further manifests this finding. Columns (1)–(3) display regression results with the block median waiting time as the dependent variable and the indicator for the London hard fork and EIP-1559 adoption rate as independent variables. The first row in Columns
(1)–(3) returns consistent and significant negative effects of the London hard fork on the waiting time, while the second row in Columns (1)–(3) returns consistent and significant negative effects of EIP-1559 adoption on waiting time. We observe that the London hard fork itself had a significant negative effect on the median waiting time of all transactions in blocks and that this effect strengthened over time. 

Specifically, the waiting time decreased on average by approximately 9 seconds after the London hard fork, and if we assume a linear treatment effect, the waiting time should further decrease by another 6-11 seconds if all users adopt EIP-1559-style bidding. The relationship between the block median waiting time, London hard fork, and EIP-1559 adoption rate is visualized in~\cref{fig:cf_wt}, which simulates the waiting time at different EIP-1559 adoption rates based on Column (3) in~\cref{table: waiting time}. 

Columns (4) and (5) in~\cref{table: waiting time} display regression results with the block median waiting time of only the legacy transactions as the dependent variable and the indicator for the London hard fork and EIP-1559 adoption rate as independent variables. Similarly, the results show that the London hard fork had a consistently and significantly negative effect on the median waiting time for legacy transactions. The same is true for the EIP-1559 adoption rate (second row in Columns (4) and (5)). However, the coefficient of these two regressions might be biased by the selection of the adoption of EIP-1559. Early adopters of EIP-1559 are likely to be more sophisticated users, such as mining pools and institutional investors. Nonetheless, the spillover effect of EIP-1559 adoption to legacy transactions is convincing given the descriptive statistics mentioned above. Additionally, the significantly positive coefficient on 90-block volatility in Column (3) implies that the waiting time is longer when Ether price volatility is higher.

\begin{table}[!htbp] \centering
\small\setlength{\tabcolsep}{0pt}
\begin{tabular}{@{\extracolsep{0.5pt}}lccccc}
\\[-1.8ex]\hline
\hline \\[-1.8ex]
& \multicolumn{5}{c}{median waiting time} \\[-1.8ex]
\cr \cline{2-6} 
& (1) & (2) & (3) & (4) & (5) \\
& All Txs & All Txs & All Txs & Legacy Txs & Legacy Txs \\
\hline \\[-1.8ex]
 London Hardfork & -9.319$^{***}$ & -9.689$^{***}$ & -8.544$^{***}$ & -4.375$^{***}$ & -3.671$^{***}$ \\
  & (0.396) & (0.396) & (0.394) & (0.405) & (0.404) \\
 EIP-1559 adoption & -6.140$^{***}$ & -5.956$^{***}$ & -11.147$^{***}$ & -21.889$^{***}$ & -26.560$^{***}$ \\
  & (0.583) & (0.583) & (0.585) & (0.597) & (0.601) \\
 nblock  & -0.000$^{***}$ & -0.000$^{***}$ & -0.000$^{***}$ & -0.000$^{*}$ & -0.000$^{***}$ \\
  & (0.000) & (0.000) & (0.000) & (0.000) & (0.000) \\
 median gas price & & -0.005$^{***}$ & 0.000$^{}$ & & -0.002$^{}$ \\
  & & (0.001) & (0.001) & & (0.001) \\
 90-block volatility & & -117.596$^{***}$ & 140.907$^{***}$ & & 133.364$^{***}$ \\
  & & (9.332) & (46.044) & & (47.266) \\
 size & & & 0.000$^{***}$ & & 0.000$^{***}$ \\
  & & & (0.000) & & (0.000) \\
 ROI & & & -232.701$^{***}$ & & -217.542$^{***}$ \\
  & & & (46.675) & & (47.913) \\
 Intercept & 28.061$^{***}$ & 24.296$^{***}$ & 15.066$^{***}$ & 28.280$^{***}$ & 16.360$^{***}$ \\
  & (0.191) & (0.381) & (0.494) & (0.196) & (0.507) \\
\hline \\[-1.8ex]
 Observations & 138,043 & 137,795 & 137,795 & 138,043 & 137,795 \\
 $R^2$ & 0.736 & 0.737 & 0.744 & 0.728 & 0.734 \\
\hline
\hline \\[-1.8ex]
\multicolumn{2}{l}{\textit{Note: Hour fixed effect included.}} & \multicolumn{4}{r}{$^{*}$p$<$0.1; $^{**}$p$<$0.05; $^{***}$p$<$0.01} \\
\end{tabular}

  \begin{tablenotes}
      \footnotesize
      \item Linear regression with the block median waiting time as the dependent variable and the indicator for the London hard fork and the EIP-1559 adoption rate as independent variables, with different sets of controls shown in different columns. Outcome variable data were trimmed to $<$300 s to avoid extreme outliers (4\% of the data). Standard errors are in parentheses. The median waiting time dropped significantly after the London hard fork. It further dropped in the blocks with a higher adoption rate of EIP-1559 style bidding. The data frequency is by block. Column (3) of this table is visualized in~\cref{Appendix D}~\cref{fig:cf_wt}.
      
      \item \textbf{Instructions for reading the tables:} 
      1) The table header shows the dependent variable (i.e., the all-transaction and legacy transaction median waiting time in each block). 2) The index column presents the independent variables. (i.e., the London hard fork, EIP-1559 adoption, nblock, etc.). 3) Each column labeled with a number presents the result of one regression in the form of ~\cref{eqn:reg} Each number presents the effect of the independent variable (the row index) on the dependent variable after we control for other independent variables (if in the same column, the output of that row is not blank); for example, in Column (1), -9.321 presents the effect of the London hard fork on the all-transaction median waiting time. 5) When the number is positive (negative), the dependent variable increases (decreases) on average by the absolute value of the number as the independent variable increases by 1 unit; for example, in Column (1), -9.321 means that when the London hard fork turns from 0 to 1 (when it happens), the transaction waiting time decreases by 9.321 on average. 6) The stars show the significance level of a t test, i.e., the error rate to reject the null hypothesis that the independent variable has no effect on the dependent variable; for example, in Column (2), there are three stars for the row with index “EIP-1559 adoption”, which means that if the null hypothesis is true (EIP-1559 adoption has no effect on the waiting time), there is only a 1\% chance that we will see data as extreme as this. 8) If a cell is left blank, the variable is not included in the regression presented by the column. With different cells left blank, we present regression results with different variables included. These instructions also guide the reading of \cref{table:size sibling} and ~\cref{table: IQR}.
  \end{tablenotes}
  \caption{Waiting Time and EIP-1559 Adoption}
  \label{table: waiting time}
\end{table}

\paragraph{Validating robustness with additional data}
To check the robustness of the observed waiting time reduction, we repeat the analysis in a sample extending to two months after the adoption of EIP-1559 . A graph similar to~\cref{fig:wtdistr} is included in the appendix (\cref{fig:wt_afterword}),but in brief, the results are consistent: when we use data running until August 27, 2021, the median waiting time decreases by 49.17\%; with data until October 21, 2021, the median waiting time decreases by 41.23\%. Both confirm that EIP-1559 significantly reduces waiting times.

\subsection{Consensus Security}
\label{sec:security}

The Ethereum community has extensively discussed the security implications of EIP-1559~\cite{buterin_2021_eip, eip_1559_spikes}, and the community has largely agreed that it should not compromise consensus security. We investigate three avenues through which EIP-1559 might affect consensus: fork rates, network load, and Miner Extractable Value (MEV). With existing evidence, we tend to believe that EIP-1559 does not make the Ethereum system substantially more insecure.

\subsubsection{Fork rates}
The fork rate is an important indicator for consensus security. The prevalence of forks (or so-called uncle blocks in Ethereum) can lead to higher vulnerability to double-spend attacks and selfish mining~\cite{gervais_2016_on}. EIP-1559 changes the distribution of block sizes, so we would like to understand its implication for fork rates. We use the terms “uncle rates” and “fork rates” interchangeably.

We investigate the relationship between EIP-1559, block gas used, block size, and the number of uncles in this section and conclude that the influence of EIP-1559 on the number of uncle blocks is marginal. We also find that EIP-1559 increased block size on average, which led to a higher fork rate. Column (1) of~\cref{table:size sibling} shows the results of a linear regression between block size and the London hard fork. This indicates that the average size increased from 64.05 kbytes (the intercept in Column (1)) before the London hard fork to 78.01 kbytes
(adding the intercept with the coefficient on the London hard fork) after the London hard fork by 13.96 kbytes (the London hard fork coefficient in Column (1)). Still using block size as the dependent variable and the London hard fork as the independent variable, Column (2) further controls for the EIP-1559 adoption rate, gas used, and the interaction term between the London hard fork and gas used. The results show that the London hard fork itself, adoption rate, and gas used are all positively and significantly associated with block size. With these two columns, we conclude that block size increased significantly after the London hard fork and is growing with EIP-1559 adoption.

Column (3) of ~\cref{table:size sibling} shows the result of logistic regression with the indicator for whether a block has siblings as the dependent variable and block size in kbytes as the independent variable. The result shows that as block size grows, the likelihood that a block has siblings becomes significantly larger. (The coefficient of block size in Column (3), 0.0034, is positive and has three asterisks.) Together with those in Columns (1) and (2), our results suggest that EIP-1559 could also increase the likelihood of sibling appearance by increasing block size. Column (4) further controls for the event of the London hard fork and EIP-1559 adoption, suggesting alternative mechanisms on how EIP-1559 might affect sibling appearance. After we control for block size, EIP-1559 still has a weak positive effect (the coefficient of the London hard fork, 0.0704, is positive with one asterisk, i.e., a larger p value) on sibling appearance, but EIP-1559 adoption (the coefficient on EIP-1559 adoption, -0.2257, is negative with three asterisks) has a significantly negative effect. These estimates present evidence that EIP-1559 affects sibling appearance mainly through block size with other unknown channels to be investigated in further research.  

\begin{table}[htbp]
\setstretch{0.9}
    \centering
\small\setlength{\tabcolsep}{1pt}
{
\begin{tabular}{@{\extracolsep{5pt}}lcc}
\\[-1.8ex]\hline
\hline \\[-1.8ex]
& \multicolumn{2}{c}{block size} \
\cr \cline{2-3}
\\[-1.8ex] & (1) & (2) \\
\hline \\[-1.8ex]
 London Hardfork & 13960.686$^{***}$ & 25126.761$^{***}$ \\
  & (225.731) & (7494.917) \\
 EIP-1559 adoption & & 25803.954$^{***}$ \\
  & & (357.959) \\
 gas used & & 0.006$^{***}$ \\
  & & (0.001) \\
 London Hardfork * gas used & & -0.002$^{***}$ \\
  & & (0.001) \\
 Intercept & 64051.066$^{***}$ & -33113.125$^{***}$ \\
  & (159.455) & (7492.033) \\
\hline \\[-1.8ex]
 Observations & 138,055 & 138,055 \\
 $R^2$ & 0.027 & 0.726 \\
\hline
\hline \\[-1.8ex]
\end{tabular}

  \begin{tablenotes}
      \footnotesize
      \item Standard errors in parentheses. Columns (1) and (2) show the linear regression with block size (in bytes) as the dependent variable and the London hard fork and EIP-1559 adoption as independent variables. Block size became larger on average after the London hard fork, and it was larger when the block had a high EIP-1559 adoption rate or high gas usage. The data frequency is by block.
  \end{tablenotes}

\begin{tabular}{@{\extracolsep{10pt}}lcc}
\\[-1.8ex]\hline
\hline \\[-1.8ex]
& \multicolumn{2}{c}{sibling indicator} \
\cr \cline{2-3}
\\[-1.8ex] & (3) & (4) \\
\hline \\[-1.8ex]
 block size (kbyte) & 0.0034$^{***}$ & 0.0036$^{***}$ \\
  & (0.0003) & (0.0003) \\
 London Hardfork & & 0.0704$^{*}$ \\
  & & (0.0400) \\
 EIP-1559 adoption & & -0.2257$^{***}$ \\
  & & (0.0764) \\
 Intercept & -3.2573$^{***}$ & -3.2608$^{***}$ \\
  & (0.0249) & (0.0263) \\
\hline \\[-1.8ex]
 Observations & 138,055 & 138,055 \\
 \multicolumn{3}{l}{Logistic regression used; $^{*}$p$<$0.1; $^{**}$p$<$0.05; $^{***}$p$<$0.01}\\
 \hline 
 \hline \\[-1.8ex]
\end{tabular}
}
  \begin{tablenotes}
      \footnotesize
      \item Standard errors in parentheses. Columns (3) and (4) show the linear regression with block size (in bytes) as the dependent variable and the London hard fork and EIP-1559 adoption as independent variables. Block size became larger on average after the London hard fork, and it was larger when the block had a high EIP-1559 adoption rate or high gas usage. The data frequency is by block.
  \end{tablenotes}

    \caption{Sibling existence, block size, and EIP-1559}
    \label{table:size sibling}
\end{table}

\subsubsection{Network load}

A debated point about the security implications of EIP-1559 is whether it will put the system under a higher load~\cite{eip_1559_spikes, buterin_2021_eip}---i.e., whether it will require nodes to perform more computational, networking, and storage work to participate in the blockchain protocol due to the increased block size cap (from 15 M to 30 M) and skewed distribution of block gas used (see~\cref{fig:gas used}).

The blog post by Buterin (2021)~\cite{eip_1559_spikes} mentions that “block variance is nothing to worry about”. One of the arguments in favor of this view is that short-term spikes happened even before the London hard fork due to the Poisson process inherent to proof-of-work mining. This being true, it is unclear how the pre–London hard fork spikes compare to the post–London hard fork ones, as EIP-1559 introduces larger blocks that might contribute to more frequent and intense spikes (the stochastic nature of block production would have the same effect before and after the London hard fork).

We define the average load of the Ethereum system in a given period $T$ as the average of gas consumed per second in $T$. We calculate the average load for varying time intervals (20--120 seconds) at each block timestamp and compare the distributions before and after the London hard fork. The results are shown in~\cref{fig:avggasper}. In the appendix, we also select various thresholds to define the load spike and calculate the percentage of load spikes before and after the London hard fork, as shown in~\Cref{Table 5.4.3}. From~\cref{fig:avggasper} and~\Cref{Table 5.4.3}, we find that EIP-1559 does not affect the integral network load or frequency of load spikes to a significant degree, especially not for an extended period (e.g., 40 seconds or 3 blocks or more). Our results confirm the argument from Buterin's blog post~\cite{eip_1559_spikes}.

\begin{figure}
\centering
\includegraphics[width=\linewidth]{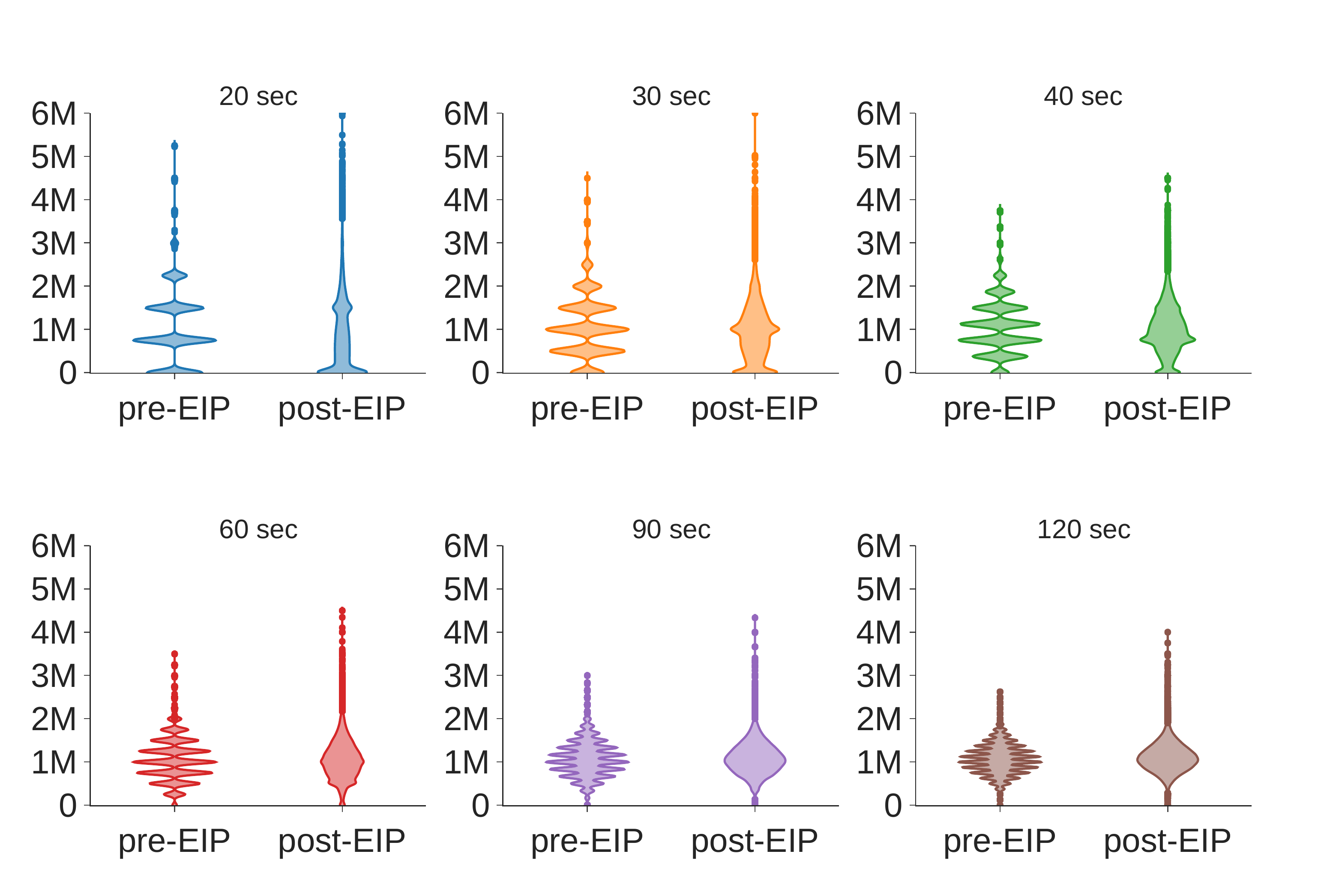}
\caption{Moving averages of block gas used per second for different time intervals}
\label{fig:avggasper}
\end{figure}

\subsubsection{Miner Extractable Value (MEV)}
MEV refers to the profit that a miner can make through her ability to arbitrarily include, exclude, or reorder transactions within the blocks she produces. 
As~\cite{daian_2019_flash} points out, significant MEV can incentivize miners to deviate from the consensus protocol (e.g., to fork or even rewind the blockchain to collect profit in MEV~\cite{daian_2019_flash}).
The volume of and changes in MEV have a profound impact on consensus.

We focus on the MEV data described in~\cref{subsec:revenue data} to observe the impact of EIP-1559 on miner revenue.
We notice that miner revenue from MEV dropped temporarily after the London fork, though it soon recovered. However, in the end, miner revenue from MEV became a much larger share of miner revenue. 
This might create an incentive for miners to invest more in MEV extraction.

\begin{figure}[!ht]\centering
  \subfigure[Source of miners' revenue]{\includegraphics[scale=.25]{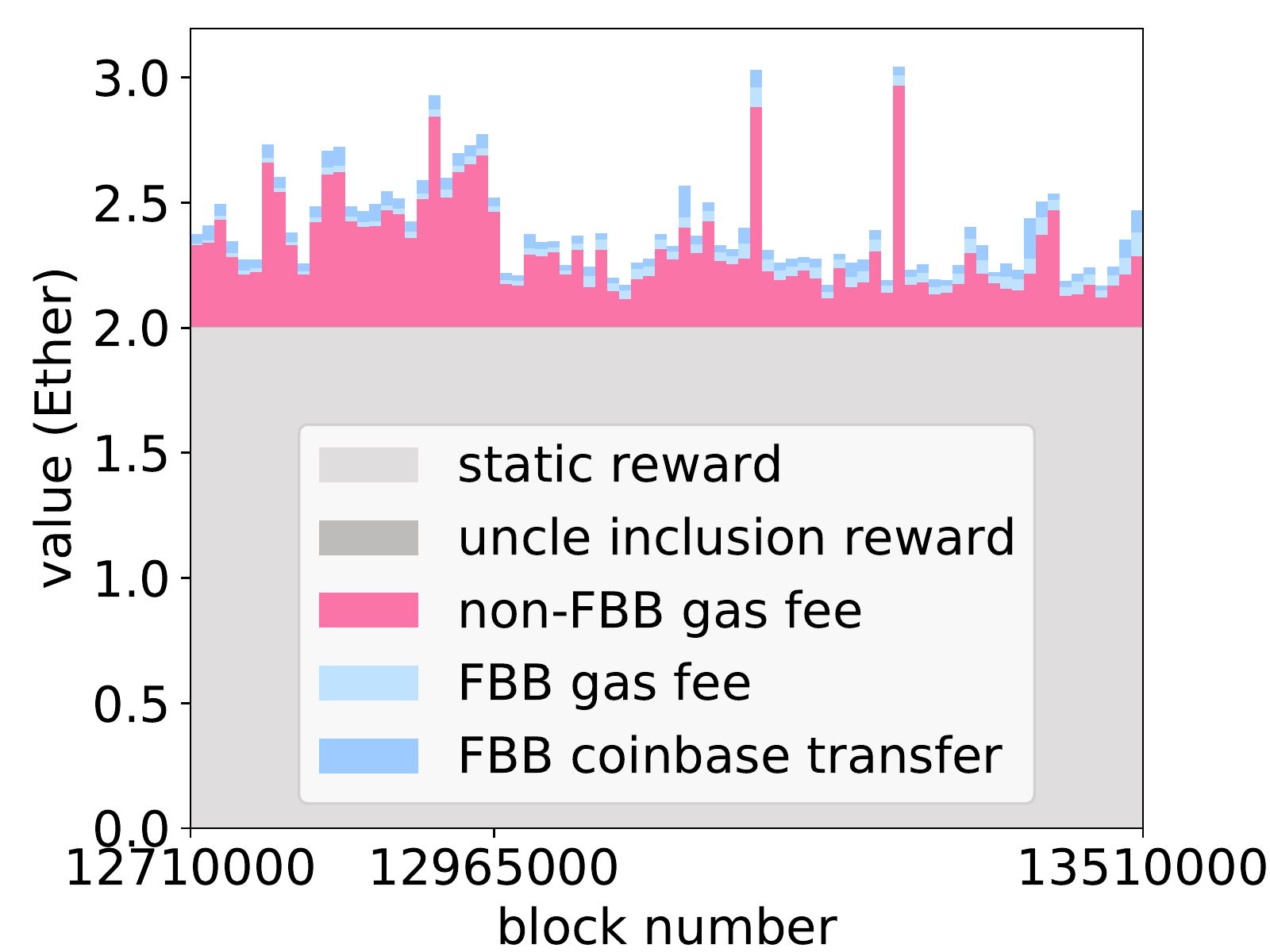}\label{MEV:whole}}
  \subfigure[MEV from Flashbots]{\includegraphics[scale=.25]{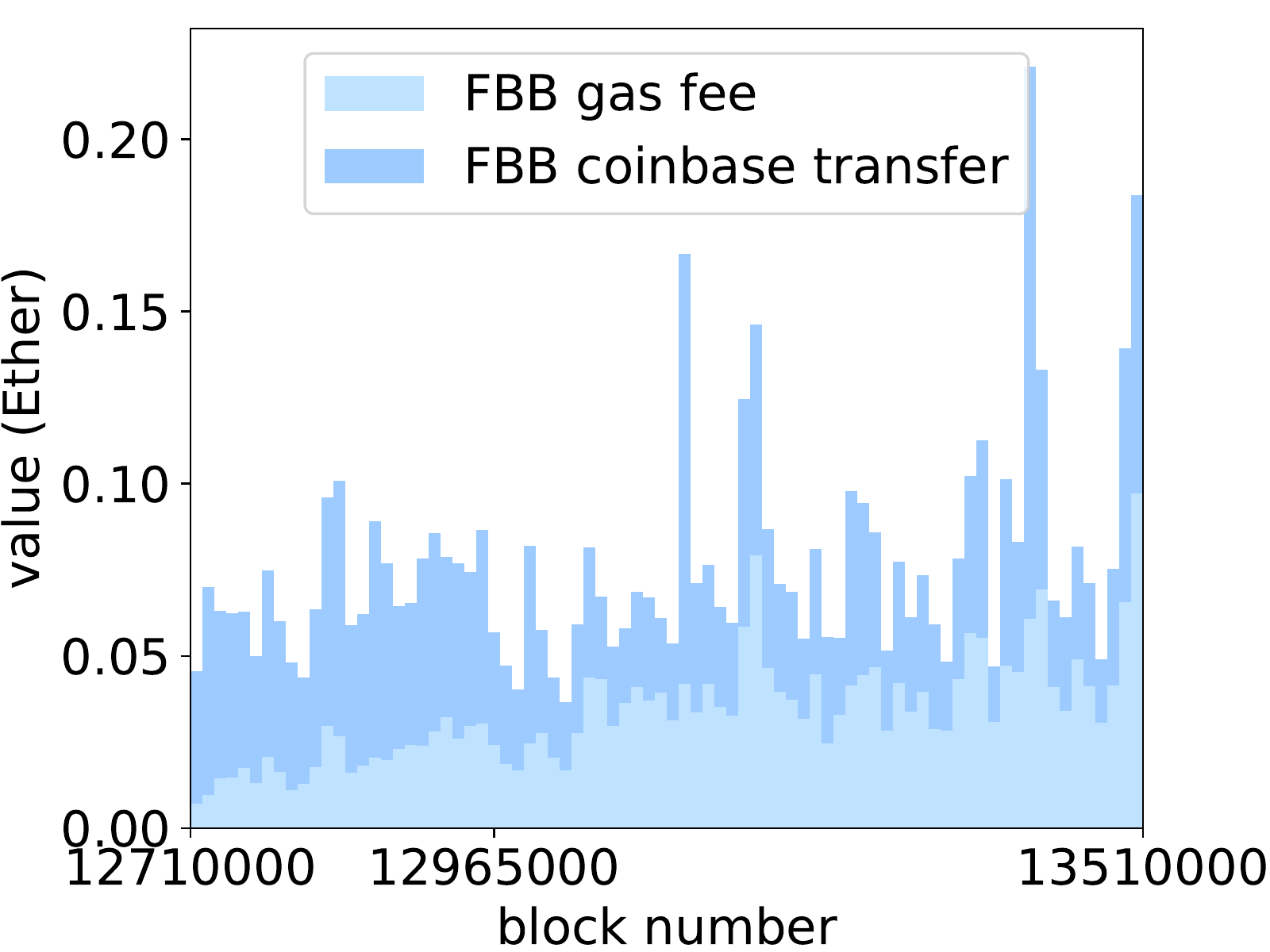}\label{MEV:flashbots}}
  \subfigure[Ratio of revenue from Flashbots to total revenue]{\includegraphics[scale=.25]{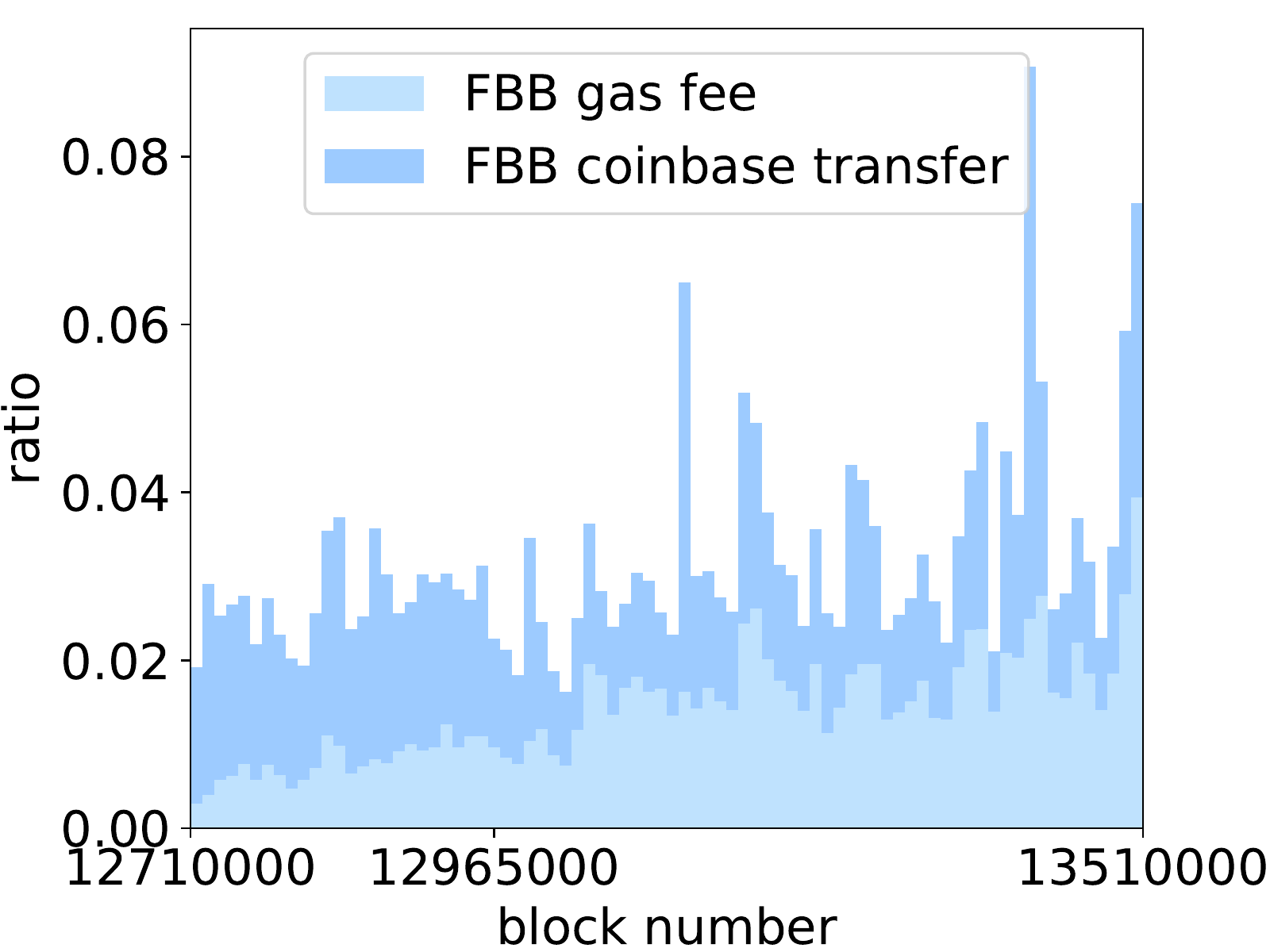}\label{MEV:ratio_total}}
  \subfigure[Ratio of revenue from Flashbots to nonstatic revenue]{\includegraphics[scale=.25]{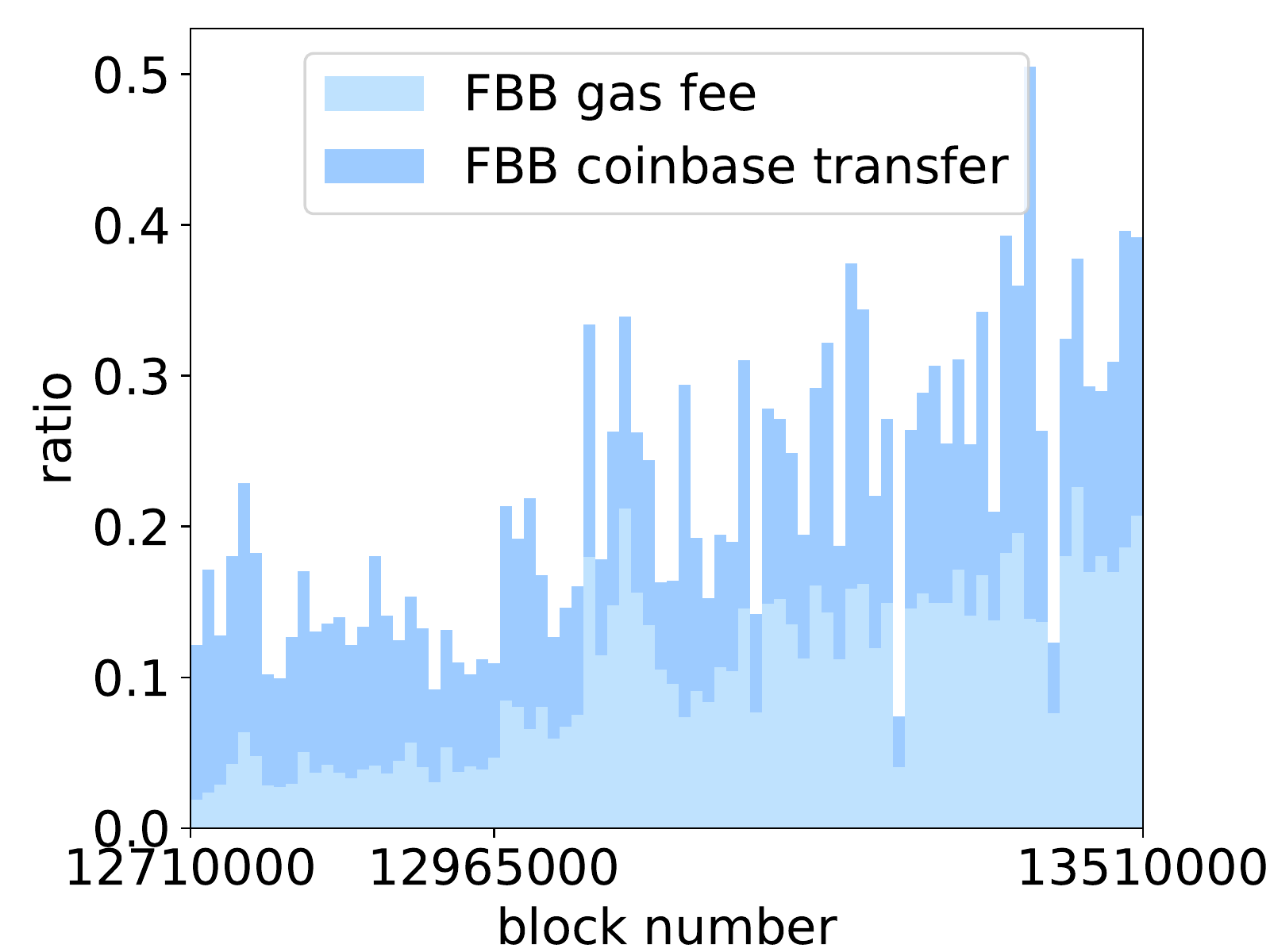}\label{MEV:ratio_nonstatic}}
  \caption{\textbf{Sources of and changes in miner revenue}}
  \label{MEV}
\end{figure}

\paragraph{Miner revenue before and after EIP-1559}
\Cref{MEV:whole} shows total miner revenue and its composition. Overall, miner revenue decreased after the EIP, primarily because the base fees are burned.

\Cref{MEV:flashbots} plots the revenue from MEV. After a downturn for less than 50,000 blocks, MEV revenue quickly recovers. This may have been for the following reasons: 1) Flashbot searchers needing to update their software after the London hard fork to adapt to EIP-1559, and 2) the potentially high volatility of miner extractable value due to network instability in the short term after the London hard fork.

\Cref{MEV:ratio_total} and~\cref{MEV:ratio_nonstatic} show the ratio of MEV revenue to total revenue and to nonstatic revenue (i.e., miner revenue minus the static block reward), respectively. As the revenue from gas fees dropped dramatically after the London hard fork while MEV revenue recovered quickly, the ratio between MEV and total revenue increased significantly. Specifically, as shown in~\cref{MEV:ratio_total} and~\cref{MEV:ratio_nonstatic}, after the London hard fork, miners' MEV revenue account for approximately 4\% of total revenue and approximately 30\% of nonstatic revenue, while before the London hard fork, the MEV revenue was only approximately 3\% of total revenue and approximately 15\% of nonstatic revenue. 

\paragraph{Distribution of nonstatic revenue before and after EIP-1559}
Miners' nonstatic revenue consists of three parts: uncle inclusion rewards, revenue from Flashbot bundles (i.e., FBB gas fees plus FBB coinbase transfers) and non-FBB gas fees.
The uncle inclusion reward is typically sm\begin{figure}[!ht]\centering
  \subfigure[Distribution of nonstatic revenue]
  {\includegraphics[scale=.25]{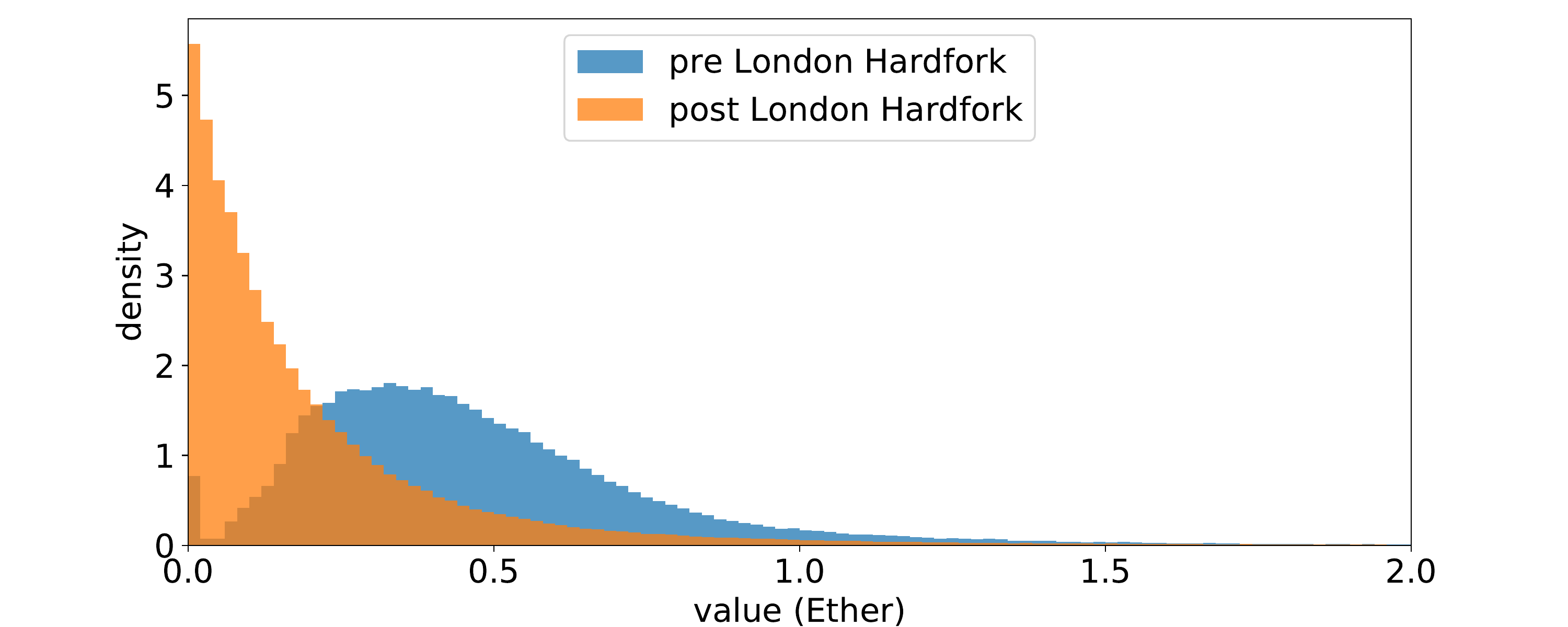}\label{MEV:dist_all}}
  \subfigure[Distribution of MEV from Flashbot bundles]
  {\includegraphics[scale=.25]{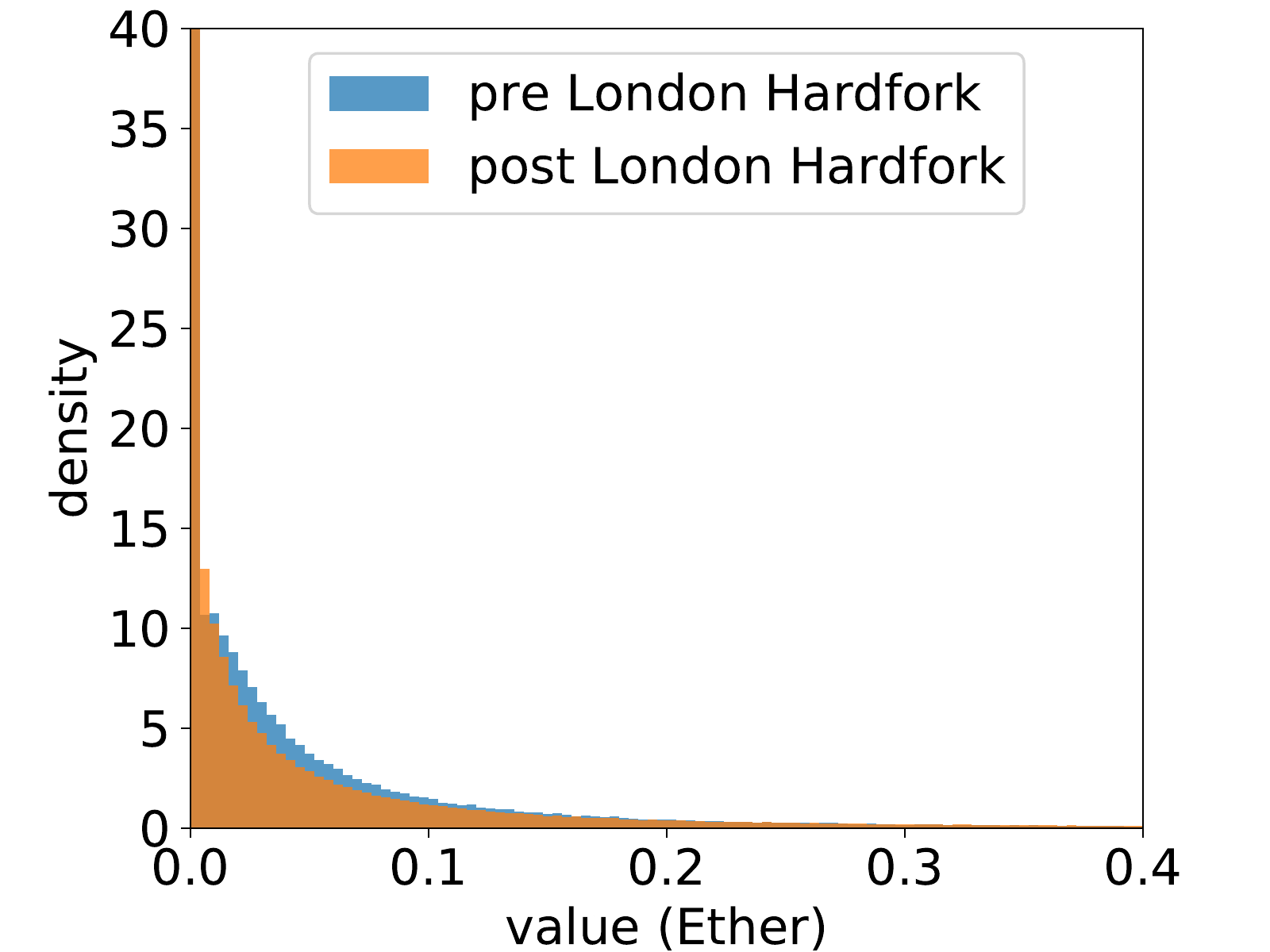}\label{MEV:dist_FBB}}
  \subfigure[Distribution of revenue from gas except from Flashbot bundles]
  {\includegraphics[scale=.25]{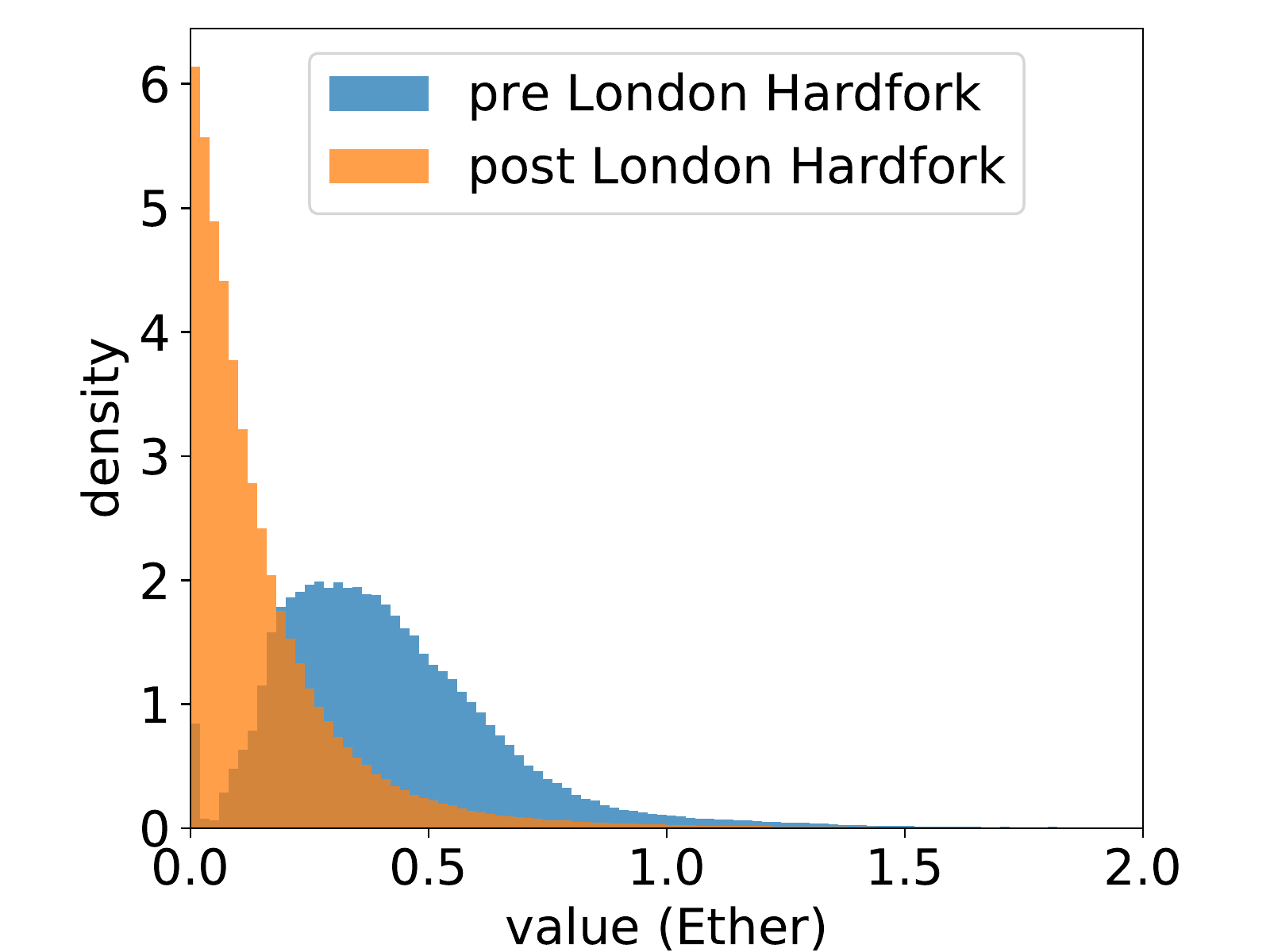}\label{MEV:dist_non}}
  \caption{\textbf{Distribution of nonstatic revenue before and after the London hard fork}}
  \label{MEV:dist}
\end{figure}
As~\cref{MEV:dist} shows, the distribution of nonstatic revenue changed significantly after EIP-1559.
\Cref{MEV:dist_FBB} and~\cref{MEV:dist_non} further break the changes down. \Cref{MEV:dist_FBB} shows the distribution of revenue from Flashbots bundles, which does not change much across the periods before and after the London hard fork.~\Cref{MEV:dist_non} shows the distribution of of non-FBB gas fees. This part of the revenue is very different in the periods before and after the London hard fork and is the main reason for the change in the distribution of miner revenue.

From~\cref{MEV:dist}, we conclude that EIP-1559 does not change the distribution of MEV revenue in the short term, but it significantly changes that of non-FBB gas fees. 
\section{Conclusion and Discussion}
\label{sec:conclusion}
We demonstrate how a major TFM reform on Ethereum affects the blockchain dynamics of Ethereum. Our empirical study relates to and tests implications for a wide range of existing theoretical research. For instance, our results on users' bids are consistent with the predictions in Roughgarden (2020) ~\cite{roughgarden_2020_transaction} that the obvious optimal bid under EIP-1559 is a max fee that represents the intrinsic value of the transaction and a max priority fee that represents the marginal cost of miners' inclusion of the transaction. Ferreira et al.~\cite{ferreira_2021_dynamic} are concerned that the EIP-1559 TFM may degrade to a first-price auction on the priority fee when the base fee is too low. Their concern is valid, but our empirical results show that this rarely happens in reality. Reijsbergen et al.~\cite{reijsbergen_2021_transaction} are concerned that volatility of base fees after EIP-1559 would make it more difficult for users to estimate transaction fees. In comparison, we show that the volatility of intrablock gas prices decreases significantly as more users adopt EIP-1559 transactions, which implies easier fee estimation and a better user experience. We also improve the strategy for measuring the transaction waiting time from previous studies, largely eliminating the influence of negative waiting time encountered by Azevedo-Sousa et al.~\cite{azevedosousa_2020_an}. 

We present several new findings that are absent from existing theoretical research. First, our results show that EIP-1559 significantly reduces waiting times and latency. However, formal modeling that could explain the effect is absent in the literature. Nonetheless, understanding waiting times in TFM is essential. Waiting times affect the user experience. Furthermore, the literature shows that latency or coexisting market congestion affects important facets of blockchain security~\cite{chung_2022_foundations, tang_2022_strategic}. The queuing theory~\cite{harcholbalter_2013_performance,li_2018_blockchain} studies how different factors, such as arrival rates, affect the waiting time of a computational system. Future research could study whether EIP-1559 affects the waiting times via the factors modeled in the queuing theory. Practically, current scalability solutions for reducing waiting times focus on layer-2 ~\cite{tsabary_2022_ledgerhedger}. Future research could consider a layer-1 solution that improves the EIP-1559 TFM. Second, we find that when Ether’s price is more volatile, the transaction fee and waiting time are significantly higher, suggesting that price volatility can be one policy parameter for the future design of TFMs. Finally, we also verify that a larger block size increases the presence of siblings. Even though the effect size through EIP-1559 is negligible, we shall consider this for future updates.

In the current research, we systematically evaluate the effects of EIP-1559, focusing on significant facets of the transaction fee, the waiting time, and consensus security. However, our methodology is generally applicable for studying the effect of EIP-1559 in particular or Ethereum Improvement Proposals in general on blockchain performances in facets such as decentralization~\cite{zhang2022sok}, network features~\cite{ao_2022_is,campajola_2022_the,decollibus_2021_heterogeneous}, and attacks~\cite{Yaish_Aviv}. For example, Cong et al. (2022)~\cite{cong_2022_inclusion} apply the same method of regression discontinuity to study EIP-1559 and find that it mitigates the identified concentration in mining rewards, token ownership, and transactions via wealth distribution. Capponi et al.(2021)~\cite{capponi_2021_proofofwork} develop a game theoretical model to study the effect of hardware efficiency and mining rewards on the decentralization level of mining. Future research can apply our methods to test whether EIP-1559 or other Ethereum improvement proposal supports decentralized cryptocurrency mining empirically. Yaish, Stern, and Zohar (2022)~\cite{cryptoeprint:2022/1020} present a risk-less attack on Ethereum’s consensus mechanism, which could also be affected by the change in TFM. Capponi et al. (2022)~\cite{capponi_2022_the} analyze the incentives for and welfare impacts of dark venue adoptions. Future research could apply our methods to study the effect of EIP-1559 on dark venue adoptions and its implications for the welfare of different stakeholders. Wang et al. (2022)~\cite{wang_2022_impact} find that non-professional users are unaware of potential financial losses due to the lack of understanding of Sandwich Attacks in the DeFi ecosystem. How does EIP-1559 impact the financial inclusion of non-professional users? We leave the question for future research. 

\section{Acknowledgments}
We benefit from conversations during the invited feature talk at ETHconomics\@ Devconnect 2022, hosted by Ethereum Foundation in Amsterdam, Netherland, presented jointly by Fan Zhang and Luyao Zhang on April 21, 2022.~\cite{Wu2022Recap} We are grateful for the insightful comments by Profs. Tim Roughgarden, Elaine Shi, Lin William Cong, Ari Juels, Ittay Eyal, Agostino Capponi, Giulia Fanti, Claudio J. Tessone, Aviv Yash, Ye Wang, Hao Chuang, and Weizhao Tang. This project is partially supported by \textit{Carry the Innovation Forward Program} jointly supported by Duke Learning Innovation at Duke University and the Center for Teaching and Learning at Duke Kunshan University. Kartik Nayak was supported in part by a VMware Early Career Grant, a Novi gift grant and grants from Ethereum Foundation. Yulin Liu thanks Bochsler Group for generously supporting his academic research. Luyao Zhang thanks Duke Kunshan University for persistently supporting her effort in interdisciplinary studies and cultivating undergraduate research. Duke Kunshan University provides the funding support for the article processing charge (APC) to facilitate the great cause of open access publication. We thank Tianyu Wu, an advisee of Luyao Zhang at Duke Kunshan University, for his assistance in checking the replicability of the data and code. Luyao Zhang, Yulin Liu, Kartik Nayak, and Fan Zhang thank Prof. Robert Calderbank, Prof. Jun Yang, and Prof. Brandon Fain for their support in Duke CS+ 2022~\cite{a2021_sciecon}, which cultivated the interdisciplinary collaboration.  

\bibliographystyle{ACM-Reference-Format}
\bibliography{main}

\pagebreak

\appendix

\section{Data Dictionary}\label{Appendix A}
We present a full view of the data sets we use and the variables in them in~\cref{table:datadict}.
\begin{table*}[!htbp]
\fontsize{7pt}{\baselineskip}\selectfont
\setlength{\tabcolsep}{1pt}
\begin{tabular}{|c|c|c|}
 \hline \hline
 Column Name & Source & Annotation \\ [1ex] 
 \hline \hline
 \multicolumn{3}{c}{\textbf{Dataset 1: Block Characteristics}} \\ [1ex] 
 \hline
 block\_number & BigQuery block data & Block number on blockchain; data key\\[0.2ex] 
 
 BQ\_timestamp & BigQuery block data & Block timestamp on blockchain\\[0.2ex] 
 
 block\_hash & BigQuery block data & Hash of the block \\[0.2ex] 
 
 sha3\_uncles & BigQuery block data & SHA3 of the uncles data in the block\\[0.2ex] 

 miner & BigQuery block data & miner address\\[0.2ex] 
 
 difficulty & BigQuery block data & Integer of the difficulty for this block\\[0.2ex] 
  
 size & BigQuery block data & The size of this block in bytes\\[0.2ex] 
 
 gas\_limit & BigQuery block data & The maximum gas allowed in this block\\[0.2ex] 
 
 gas\_used & BigQuery block data & The total used gas by all transactions in this block\\[0.2ex] 
 
 tr\_count & BigQuery block data & Number of all transactions included in this block\\[0.2ex] 
 
 base\_fee & BigQuery block data & Protocol base fee per gas \\[0.2ex] 
 
 \hline
  \multicolumn{3}{c}{\textbf{Dataset 2: Fee Dynamics}} \\ [1ex] 
 \hline 
 
 block\_number & BigQuery(BQ) transaction(tx) data & Block number on blockchain; data key\\[0.2ex] 
 
 all\_gpq[x] & BQ tx data (aggregated on block level) & [x]\% quartile of actual gas price paid per gas for all transactions in this block  \\[0.2ex] 
 
 all\_gpcount & BQ tx data (aggregated on block level) & number of transactions with valid gas price paid in this block (= tr\_count) \\[0.2ex] 
 
 all\_prq[x] & BQ tx data (aggregated on block level) & [x]\% quartile of priority fee per gas bid by all transactions in this block  \\[0.2ex] 
 
 all\_prcount & BQ tx data (aggregated on block level) & number of transactions with valid priority fee bid in this block (= number of txs that adopt EIP1559 style bid) \\[0.2ex] 
 
 all\_mfq[x] & BQ tx data (aggregated on block level) & [x]\% quartile of max fee per gas bid by all transactions in this block  \\[0.2ex] 
 
 all\_mfcount & BQ tx data (aggregated on block level) & number of txs that adopt EIP1559 style bid \\[0.2ex] 
 
 
 legacy\_gpq[x] & BQ tx data (aggregated on block level) & [x]\% quartile of actual gas price paid per gas for all transactions that adopts legacy style bid in this block  \\[0.2ex] 
 
 eip\_gpq[x]  & BQ tx data (aggregated on block level) & [x]\% quartile of actual gas price paid per gas for all transactions that adopts EIP1559 style bid in this block  \\[0.2ex] 
 
 \hline
  \multicolumn{3}{c}{\textbf{Dataset 3: Waiting Time}} \\ [1ex] 
 \hline  
 
 block\_number & Go Ethereum (Geth) & Block number on blockchain; data key\\[0.2ex] 
 
 uncle\_cnt & Geth  & Number of uncles of this block \\[0.2ex] 
 
 all\_cnt & Geth  & Number of all transactions included in this block \\[0.2ex]  
 
 all\_latetx\_cnt & Geth & Number of late transactions in this block \\[0.2ex] 
 
 all\_nevertx\_cnt & Geth & Number of never transactions in this block \\[0.2ex] 
 
 all\_wtq[x] & Geth & [x]\% quartile of waiting time for all transactions included in this block \\[0.2ex] 
 
 legacy\_cnt & Geth  & Number of transactions that adopt legacy bid style included in this block \\[0.2ex] 
 
 legacy\_latetx\_cnt & Geth & Number of late transactions that adopt legacy bid style in this block \\[0.2ex] 
 
 legacy\_nevertx\_cnt & Geth & Number of never transactions that adopt legacy bid style in this block \\[0.2ex] 
 
 legacy\_wtq[x] & Geth & [x]\% quartile of waiting time for transactions that adopt legacy bid style included in this block \\[0.2ex] 
 
 eip\_cnt & Geth  & Number of transactions that adopt EIP-1559 bid style included in this block \\[0.2ex] 
 
 eip\_latetx\_cnt & Geth & Number of late transactions that adopt EIP-1559 bid style in this block \\[0.2ex] 
 
 eip\_nevertx\_cnt & Geth & Number of never transactions that adopt EIP-1559 bid style in this block \\[0.2ex] 
 
 eip\_wtq[x] & Geth & [x]\% quartile of waiting time for transactions that adopt EIP-1559 bid style included in this block \\[0.2ex] 
 
 \hline
 \multicolumn{3}{c}{\textbf{Dataset 4: Miner Extractable Value (MEV)}} \\ [1ex] 
 \hline  
 
 block\_number & Geth & Block number on blockchain; data key\\[0.2ex] 
 
 FBB\_coinbase \_transfer & Flashbots API & Sum of coinbase transfers of all Flashbot bundles (FBB) transactions in this block \\[0.2ex] 
 
 FBB\_gas\_fee & Flashbots API & Sum of total gas fees paid by all FBB transactions in this block \\[0.2ex] 
 
 non\_FBB\_gas\_fee & Geth and Flashbots API & Sum of total gas fees paid to miners (not including burnt fees) by all non-FBB transactions in this block \\[0.2ex] 

 static\_reward  & Geth & Static reward to miners in this block \\[0.2ex] 
 
 uncle\_incl\_reward & Geth & Uncle inclusion reward to miners in this block\\[0.2ex] 
 \hline \hline 
\end{tabular}
\caption{Data dictionary}
\label{table:datadict}
\end{table*}

\section{Statistics Tests}
\begin{figure}[!htbp]
      \centering
      \includegraphics[width = \linewidth]{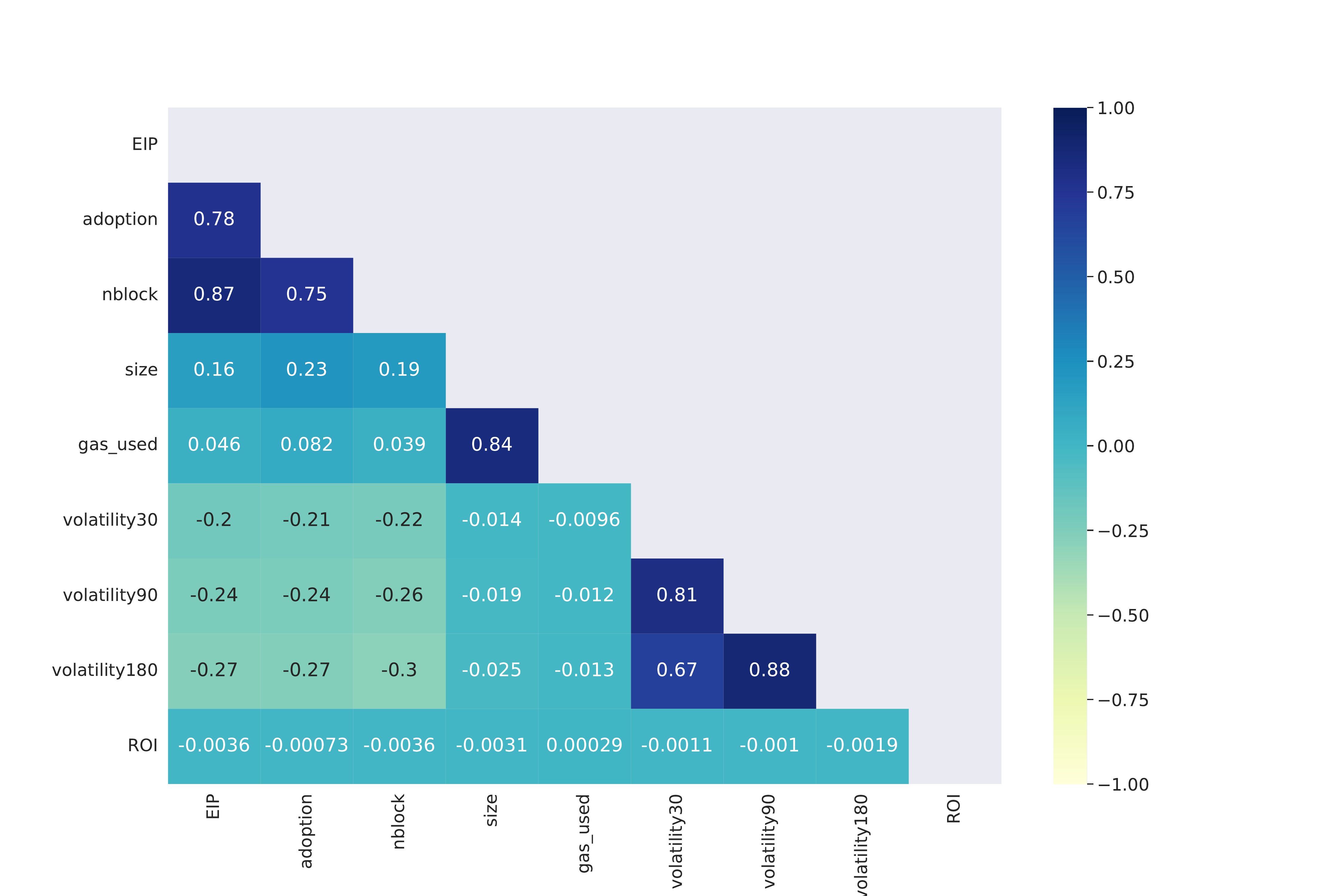}
      \caption{Correlation Heatmap for Control Variables}
      \label{fig:correlation}
  \end{figure}

\paragraph{Correlation Tests}
\Cref{fig:correlation} presents the triangle correlation heatmap between the control variables we use.
  
\paragraph{Stationary Tests}\label{Appendix B}
\Cref{table:stationtest} presents the results of the Augmented Dickey-Fuller (ADF) test for all variables we use. It proves that they are all stationary.
  \begin{table}[!htbp] \centering
\small
\setlength{\tabcolsep}{5pt}
\begin{tabular}{@{\extracolsep{5pt}}lccc}
\\[-1.8ex]\hline
\hline \\[-1.8ex]
   Variable Name  & ADF Statistics & p-value & Stationary \\
\hline \\[-1.8ex]
  base fee & -25.65 & 0.000 & Yes \\
  EIP-1559 adoption & -7.58  & 0.000 & Yes \\
  median gas price & -32.92 & 0.000 & Yes \\
  IQR & -29.24 & 0.000 & Yes \\
  standardized IQR & -34.84 & 0.000 & Yes \\
  median waiting time & -60.83 & 0.000 & Yes \\
  90min MA volatility & -10.99 & 0.000 & Yes \\
  block size & -24.50 & 0.000 & Yes \\
  block gas used & -46.85 & 0.000 & Yes \\
 \hline
\hline \\[-1.8ex]
\end{tabular}
  \caption{Dickey-Fuller Test for Stationary Variables}
  \label{table:stationtest}
\end{table}

\begin{figure}[!htbp]
      \centering
      \includegraphics[width = \linewidth]{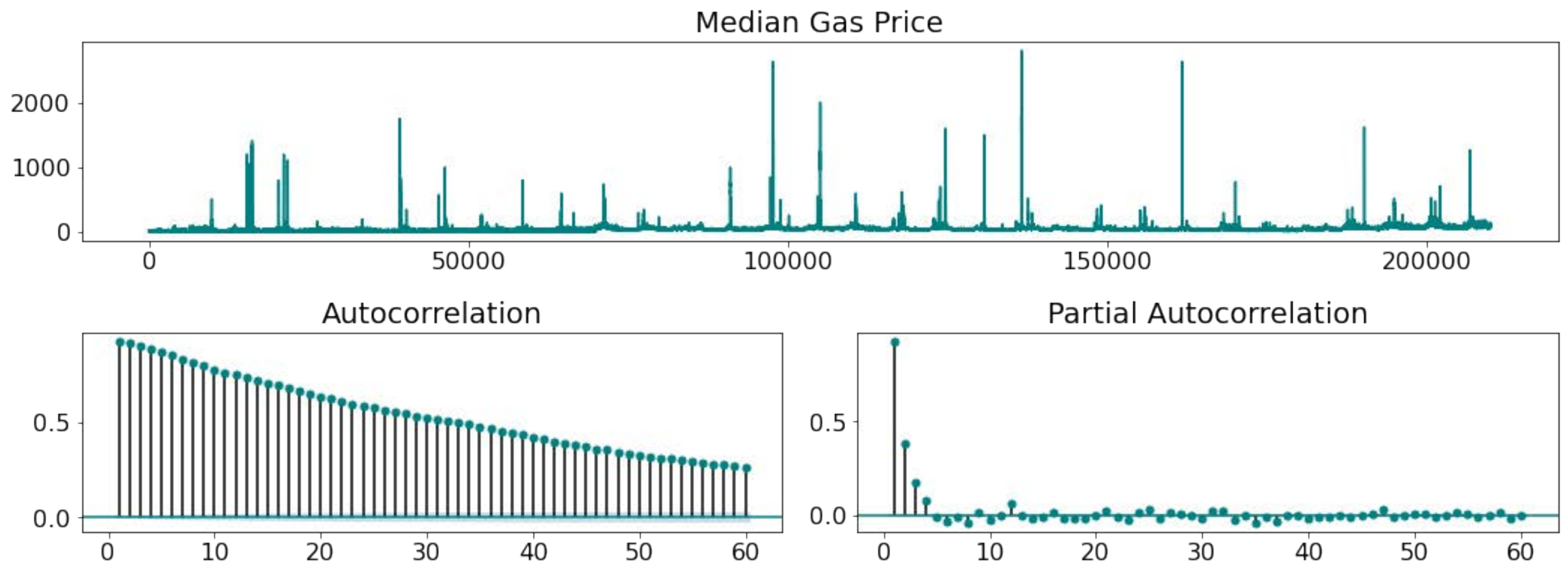}
      \vskip 12pt
      \includegraphics[width = \linewidth]{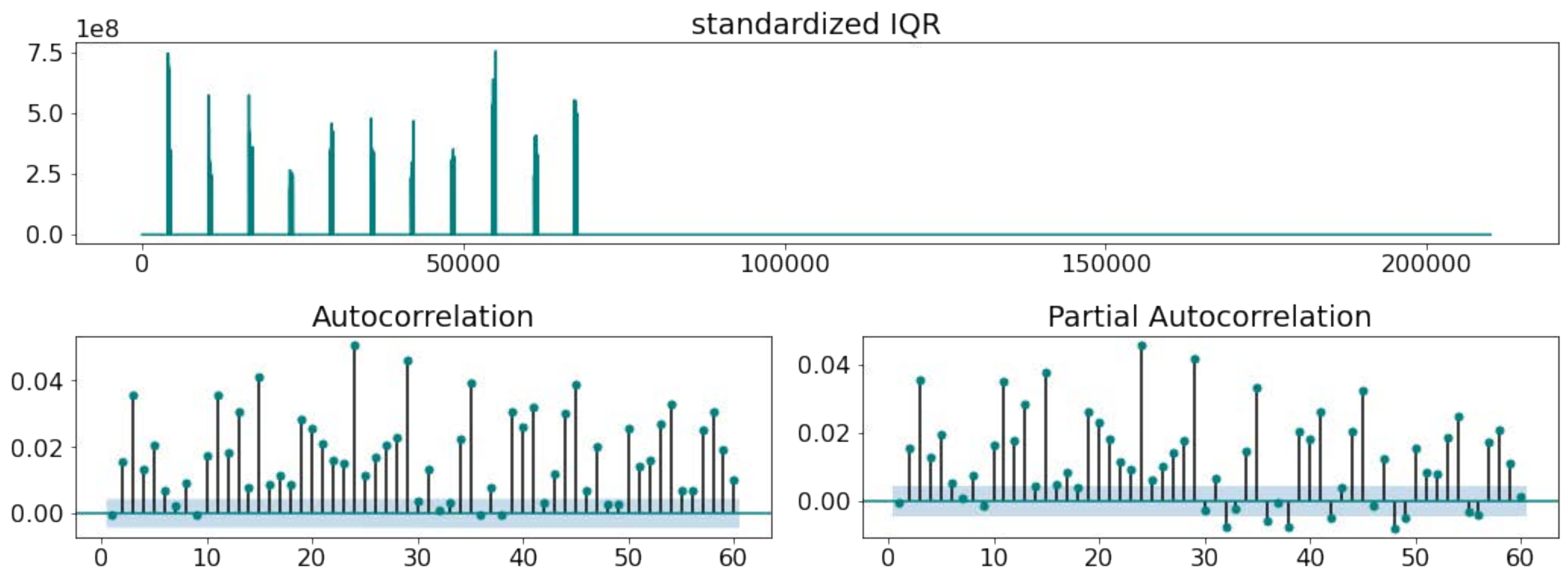}
      \vskip 12pt
      \includegraphics[width = \linewidth]{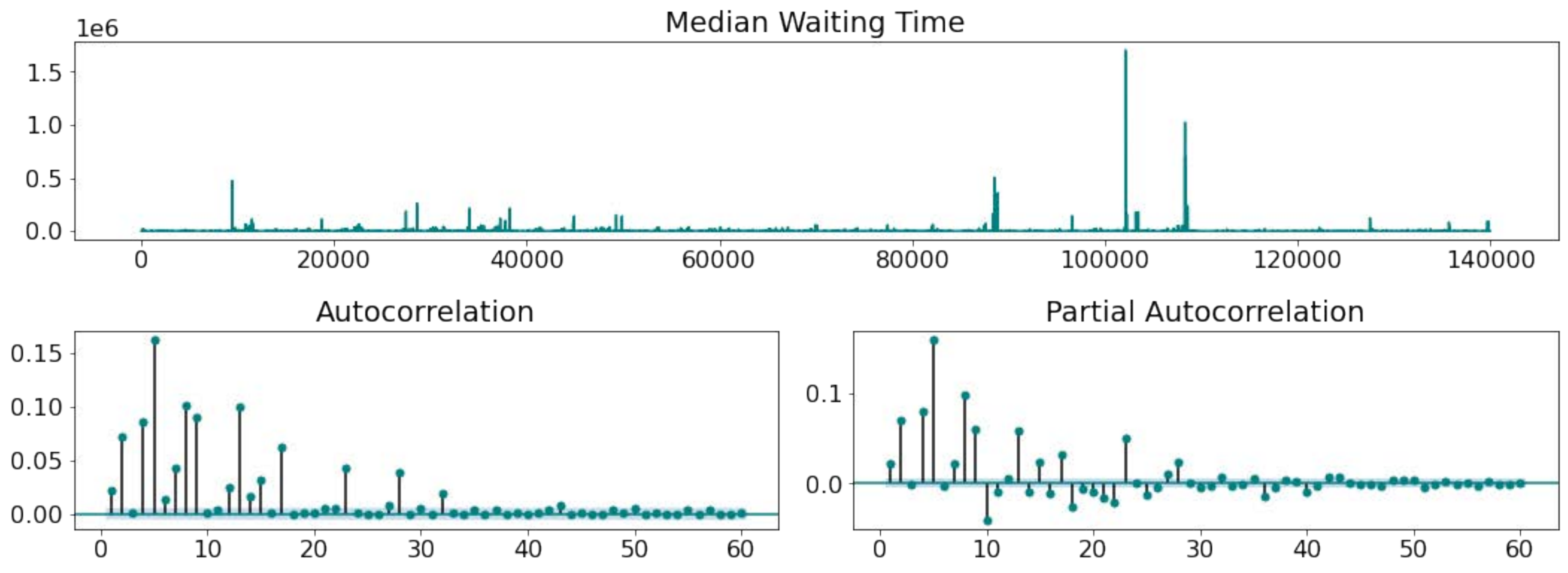}
      
      \caption{Auto-correlation of median gas price, standardized IQR, and median waiting time}
      \label{fig:AC}
  \end{figure}

\paragraph{Auto-correlation Tests}
\Cref{fig:AC} presents the auto-correlation test for the dependent variables we put in the regressions. Gas prices have a significant auto-correlation, while the autocorrelation of standardized IQR and waiting time are trivial.

\section{Robustness Checks}\label{Appendix C}
\paragraph{Fee Dynamics}
\Cref{table:C1} shows that though median gas prices increased overall after London hard fork without controls, they dropped if controlling time trend and price volatility.
  \begin{table}[htbp] \centering
  \small
\setlength{\tabcolsep}{2pt}
\begin{tabular}{@{\extracolsep{5pt}}lccc}
\\[-1.8ex]\hline
\hline \\[-1.8ex]
& \multicolumn{3}{c}{\textit{Median Gas Price}} \
\cr \cline{2-4}
\\[-1.8ex] & (1) & (2) & (3) \\
\hline \\[-1.8ex]
London Hardfork & 19.9332$^{***}$ & -12.2320$^{***}$ & -14.1355$^{***}$ \\
  & (0.3407) & (0.7284) & (0.7267) \\
EIP-1559 adoption & & -3.4519$^{***}$ & 3.1717$^{***}$ \\
  & & (1.0429) & (1.0508) \\
 & & (1.0776) & (1.0821) \\
 nblock & & 0.0005$^{***}$ & 0.0005$^{***}$ \\
  & & (0.0000) & (0.0000) \\
block size & & & -0.0001$^{***}$ \\
  & & & (0.0000) \\
 ROI & & & -68.5522$^{}$ \\
  & & & (332.4094) \\
90-block volatility & & & 1832.0280$^{***}$ \\
  & & & (91.8924) \\
 Intercept & 45.2454$^{***}$ & 61.6240$^{***}$ & 58.9252$^{***}$ \\
  & (0.8539) & (0.8952) & (1.1433) \\
\hline \\[-1.8ex]
 Observations & 138,043 & 138,043 & 137,770 \\
 $R^2$ & 0.0898 & 0.1106 & 0.1207 \\
\hline
\hline \\[-1.8ex]
\textit{Note: Hour fiex effect included} & \multicolumn{3}{r}{$^{*}$p$<$0.1; $^{**}$p$<$0.05; $^{***}$p$<$0.01} \\
\end{tabular}
  \begin{tablenotes}
      \footnotesize
      \item Linear regression with block median gas price as dependent variable, indicator of London Hardfork and EIP-1559 adoption rate as independent variables with different sets of controls shown in different columns.  Standard errors are in parentheses. Results are mixed: median gas prices increased overall after London Hardfork without controls, but they in fact dropped if controlling time trend and price volatility.
  \end{tablenotes}
  \caption{Median Gas Price and EIP-1559 adoption}
  \label{table:C1}
\end{table}
   
~\Cref{table: IQR} displays the regression outcome with standardized IQR as the dependent variable, the indicator variable of London hard fork and EIP-1559 adoption rate as independent variables. Column (1) displays that the aggregate effect of the network upgrade to the intrablock price variance is negative. Columns (2) - (4), carrying an expanding set of control variables, return a consistent estimate on the immediate and short-term effect of EIP-1559 adoption. As shown, the coefficients in the first row on London hard fork in Columns (2) - (4) are consistently and significantly positive. In contrast, the coefficients in the second row on EIP-1559 adoption are similarly and significantly negative.
   \begin{table}[!htbp] \centering\small\setlength{\tabcolsep}{1pt}
{
\begin{tabular}{@{\extracolsep{5pt}}lcccc}
\\[-1.8ex]\hline
\hline \\[-1.8ex]
& \multicolumn{4}{c}{standardized inter-quartile range} \
\cr \cline{2-5}
\\[-1.8ex] & (1) & (2) & (3) & (4) \\
\hline \\[-1.8ex]
 London Hardfork & -0.02499$^{***}$ & 0.08441$^{***}$ & 0.08425$^{***}$ & 0.08599$^{***}$ \\
  & (0.00123) & (0.00260) & (0.00260) & (0.00260) \\
 EIP-1559 adoption & & -0.23610$^{***}$ & -0.23871$^{***}$ & -0.24552$^{***}$ \\
  & & (0.00370) & (0.00370) & (0.00374) \\
 nblock & & -0.00000$^{***}$ & -0.00000$^{***}$ & -0.00000$^{***}$ \\
  & & (0.00000) & (0.00000) & (0.00000) \\
 median gas price & & & -0.00002$^{**}$ & -0.00001$^{}$ \\
  & & & (0.00001) & (0.00001) \\
 90-block volatility & & & 1.20477$^{***}$ & -1.64872$^{***}$ \\
  & & & (0.06545) & (0.31787) \\
 block size & & & & 0.00000$^{***}$ \\
  & & & & (0.00000) \\
 ROI & & & & 3.04883$^{***}$ \\
  & & & & (0.33301) \\
 Intercept & 0.21290$^{***}$ & 0.20166$^{***}$ & 0.24646$^{***}$ & 0.26316$^{***}$ \\
  & (0.00090) & (0.00132) & (0.00278) & (0.00406) \\
\hline \\[-1.8ex]
 Observations & 136,111 & 136,111 & 135,850 & 135,850 \\
 $R^2$ & 0.12276 & 0.15133 & 0.15385 & 0.15471 \\

\hline \\[-1.8ex]
\hline \\[-1.8ex]
\multicolumn{2}{l}{\textit{Note: Hour fixed effect included.}} & \multicolumn{3}{r}{$^{*}$p$<$0.1; $^{**}$p$<$0.05; $^{***}$p$<$0.01} \\
\end{tabular}
}
  \begin{tablenotes}
      \footnotesize
      \item Linear regression with standardized IQR as dependent variable, indicator of London Hardfork and EIP-1559 adoption rate as independent variables, with different sets of controls shown in different columns. Outcome variable data deleted if Q50 gas price < 1 to avoid divided by zero error (standardized IQR = IQR/Q50 gas price). Standard errors are in parentheses. Though standardized IQR increased immediately after London Hardfork, it dropped significantly and in greater scale in the blocks with more transactions adopting EIP-1559 bidding. The long term negative effect outweighs a small short term positive effect. The data frequency is by block. Column (4) is visualized in~\cref{fig:cf_iqdiff}.
  \end{tablenotes}
  \caption{Standardized IQR and EIP-1559 Adoption}
  \label{table: IQR}
\end{table}

\Cref{table:C2} shows that though IQR increased immediately after London hard fork, it dropped significantly and in greater scales in the blocks with more transactions adopting EIP-1559 bidding.
  \begin{table}[htbp] \centering
\small
\setlength{\tabcolsep}{1pt}
\begin{tabular}{@{\extracolsep{2pt}}lcccc}
\\[-1.8ex]\hline
\hline \\[-1.8ex]
& \multicolumn{4}{c}{\textit{Gas Price IQR}} \
\cr \cline{2-5}
\\[-1.8ex] & (1) & (2) & (3) & (4) \\
\hline \\[-1.8ex]
 London Hardfork & 3.47663$^{***}$ & 2.21086$^{***}$ & 2.97705$^{***}$ & 3.03820$^{***}$ \\
  & (0.05741) & (0.11904) & (0.11386) & (0.11414) \\
 EIP-1559 adoption & & -15.76611$^{***}$ & -15.11516$^{***}$ & -15.27405$^{***}$ \\
  & & (0.17048) & (0.16327) & (0.16468) \\
 nblock & & 0.00011$^{***}$ & 0.00008$^{***}$ & 0.00008$^{***}$ \\
  & & (0.00000) & (0.00000) & (0.00000) \\
 median gas price & & & 0.07002$^{***}$ & 0.07041$^{***}$ \\
  & & & (0.00063) & (0.00063) \\
 90-block volatility & & & 313.02075$^{***}$ & 308.75973$^{***}$ \\
  & & & (14.45865) & (14.46814) \\
 block size & & & & 0.00000$^{***}$ \\
  & & & & (0.00000) \\
 ROI & & & & 99.00057$^{*}$ \\
  & & & & (52.16876) \\
 Intercept & 8.72227$^{***}$ & 12.71996$^{***}$ & 6.67991$^{***}$ & 6.35035$^{***}$ \\
  & (0.14461) & (0.14694) & (0.17619) & (0.18184) \\
\hline \\[-1.8ex]
 Observations & 137,271 & 137,271 & 136,998 & 136,998 \\
 $R^2$ & 0.05731 & 0.13193 & 0.21016 & 0.21048 \\

\hline \\[-1.8ex]
\multicolumn{2}{l}{\textit{Note: Hour fixed effect included.}} & \multicolumn{3}{r}{$^{*}$p$<$0.1; $^{**}$p$<$0.05; $^{***}$p$<$0.01} \\
\end{tabular}
  \begin{tablenotes}
      \footnotesize
      \item Linear regression with gas price IQR as dependent variable, indicator of London Hardfork and EIP-1559 adoption rate as independent variables, with different sets of controls shown in different columns. Standard errors are in parentheses. Though IQR increased immediately after London Hardfork, it dropped significantly and in greater scales in the blocks with more transactions adopting EIP-1559 bidding. The long term negative effect outweighs a small short term positive effect.
  \end{tablenotes}
  \caption{Intra-block Gas Price Difference and EIP-1559 Adoption - Measured by IQR}
  \label{table:C2}
\end{table}
  
\Cref{table:C3} shows that our results on standardized IQR are robust to different volatility measures.
  \begin{table}[htbp] \centering
\small
\setlength{\tabcolsep}{4pt}
\begin{tabular}{@{\extracolsep{5pt}}lccc}
\\[-1.8ex]\hline
\hline \\[-1.8ex]
& \multicolumn{3}{c}{\textit{standardized IQR}} \
\cr \cline{2-4}
\\[-1.8ex] & (1) & (2) & (3) \\
\hline \\[-1.8ex]
London Hardfork & 0.08620$^{***}$ & 0.08602$^{***}$ & 0.08588$^{***}$ \\
  & (0.00260) & (0.00260) & (0.00260) \\
EIP-1559 adoption & -0.24701$^{***}$ & -0.24567$^{***}$ & -0.24466$^{***}$ \\
  & (0.00374) & (0.00374) & (0.00375) \\
nblock & -0.00000$^{***}$ & -0.00000$^{***}$ & -0.00000$^{***}$ \\
  & (0.00000) & (0.00000) & (0.00000) \\
median gas price& -0.00001$^{}$ & -0.00001$^{}$ & -0.00001$^{}$ \\
  & (0.00001) & (0.00001) & (0.00001) \\
30-block volatility & -4.13270$^{***}$ & & \\
  & (0.38052) & & \\
90-block volatility & & -1.77087$^{***}$ & \\
  & & (0.33064) & \\
180-block volatility & -0.48957$^{**}$ \\
  & & & (0.21862) \\
block size & 0.00000$^{***}$ & 0.00000$^{***}$ & 0.00000$^{***}$ \\
  & (0.00000) & (0.00000) & (0.00000) \\
 ROI & 1.44063$^{}$ & 1.45541$^{}$ & 1.46281$^{}$ \\
  & (1.19449) & (1.19488) & (1.19499) \\
 Intercept & 0.27425$^{***}$ & 0.26423$^{***}$ & 0.25733$^{***}$ \\
  & (0.00400) & (0.00414) & (0.00421) \\
\hline \\[-1.8ex]
 Observations & 135,730 & 135,730 & 135,730 \\
$R^2$ & 0.04572 & 0.04510 & 0.04493 \\
\hline
\hline \\[-1.8ex]
\multicolumn{2}{l}{\textit{Note: Hour fixed effect included.}} & \multicolumn{2}{r}{$^{*}$p$<$0.1; $^{**}$p$<$0.05; $^{***}$p$<$0.01} \\
\end{tabular}
  \begin{tablenotes}
      \footnotesize
      \item Linear regression with standardized IQR as dependent variable, indicator of London Hardfork and EIP-1559 adoption rate as independent variables, with different price volatility measures shown in different columns. Standard errors are in parentheses. Our results are robust to different volatility measures.
  \end{tablenotes}

  \caption{Intra-block Gas Price Difference and EIP-1559 Adoption - Different Volatility Measures}
  \label{table:C3}
\end{table}

\paragraph{Waiting Time}
  
\Cref{table:C4} shows that least absolute deviation (LAD) regression results are similar to OLS regressions in~\cref{table: waiting time}, but the coefficient estimates are overall smaller in scale.
  \begin{table}[htbp] \centering
\small
\setlength{\tabcolsep}{1pt}
\begin{tabular}{@{\extracolsep{5pt}}lcccc}
\\[-1.8ex]\hline
\hline \\[-1.8ex]
& \multicolumn{4}{c}{\textit{Median Waiting Time}} \
\cr \cline{2-5}
\\[-1.8ex] & (1) & (2) & (3) & (4) \\
\hline \\[-1.8ex]
London Hardfork & -6.07300$^{***}$ & -4.95758$^{***}$ & -4.99889$^{***}$ & -2.90102$^{***}$ \\
  & (0.07491) & (0.16097) & (0.16210) & (0.09777) \\
EIP-1559 adoption & & -0.24513$^{}$ & 0.05707$^{}$ & -4.52863$^{***}$ \\
  & & (0.23778) & (0.23920) & (0.14574) \\
 nblock & & -0.00001$^{***}$ & -0.00001$^{***}$ & -0.00003$^{***}$ \\
  & & (0.00000) & (0.00000) & (0.00000) \\
median gas price & & & -0.00151$^{**}$ & 0.00805$^{***}$ \\
  & & & (0.00060) & (0.00036) \\
90-block volatility & & & 212.95781$^{***}$ & -0.06064$^{***}$ \\
  & & & (19.51700) & (0.00051) \\
block size & & & & 0.00012$^{***}$ \\
  & & & & (0.00000) \\
 ROI & & & & -0.00249$^{***}$ \\
  & & & & (0.00002) \\
 Intercept & 13.14800$^{***}$ & 12.82830$^{***}$ & 11.40213$^{***}$ & 5.20402$^{***}$ \\
  & (0.16556) & (0.17339) & (0.20453) & (0.11094) \\
\hline \\[-1.8ex]
 Observations & 140,000 & 138,043 & 137,777 & 137,777 \\
\hline
\hline \\[-1.8ex]
\multicolumn{2}{l}{\textit{Note: Hour fixed effect included.}} & \multicolumn{3}{r}{$^{*}$p$<$0.1; $^{**}$p$<$0.05; $^{***}$p$<$0.01} \\
\end{tabular}
  \begin{tablenotes}
      \footnotesize
      \item Least absolute deviation (LAD) regression with block median waiting time as dependent variable, indicator of London Hardfork and EIP-1559 adoption rate as independent variables with different sets of controls shown in different columns. Standard errors are in parentheses. Results are similar to OLS regressions in~\cref{table: waiting time}, but the coefficient estimates are overall smaller in scale.
  \end{tablenotes}
  \caption{Ethereum Median Waiting Time and EIP-1559 Adoption - LAD Regression}
  \label{table:C4}
\end{table}
  
Column (3) in \Cref{table:C5} shows that transaction type had trivial effects on median waiting time.
  \begin{table}[htbp] \centering
\small
\setlength{\tabcolsep}{3pt}
\begin{tabular}{@{\extracolsep{5pt}}lccc}
\\[-1.8ex]\hline
\hline \\[-1.8ex]
& \multicolumn{3}{c}{\textit{Median Waiting Time}} \
\cr \cline{2-4}
\\[-1.8ex] & (1) & (2) & (3) \\
\hline \\[-1.8ex]
 TxType == 2 & -0.49933$^{***}$ & -1.14876$^{***}$ & 0.00059$^{}$ \\
  & (0.13514) & (0.12950) & (0.00088) \\
 nblock & 0.00005$^{***}$ & 0.00001$^{**}$ & \\
  & (0.00000) & (0.00000) & \\
 EIP-1559 adoption & -3.83371$^{***}$ & -9.74894$^{***}$ & \\
  & (0.30980) & (0.30025) & \\
90-block volatility & & 59.16974$^{}$ & \\
  & & (37.05756) & \\
block size & & 0.00013$^{***}$ & \\
  & & (0.00000) & \\
 ROI & & -125.41390$^{***}$ & \\
  & & (36.26787) & \\
 Intercept & 14.76364$^{***}$ & 5.37660$^{***}$ & -0.00345$^{}$ \\
  & (0.15423) & (0.30356) & (0.03561) \\
\hline \\[-1.8ex]
 Observations & 137,780 & 137,780 & 140,000 \\
 $R^2$ & 0.70541 & 0.73170 & 0.00000 \\

\hline
\hline \\[-1.8ex]
 & \multicolumn{3}{r}{$^{*}$p$<$0.1; $^{**}$p$<$0.05; $^{***}$p$<$0.01} \\
\end{tabular}
  \begin{tablenotes}
      \footnotesize
      \item Fixed effect regression with median waiting as dependent variable, indicator of transaction type as independent variable, with different sets of controls shown in different columns. Hour fixed effect included in Column (1) and (2), while block fixed effect included in Column (3). Standard errors are in parentheses. Column (3) shows that transaction type had trivial effects on median waiting time.
  \end{tablenotes}
  \caption{Comparison of Waiting Time between Legacy Style Transactions and EIP-1559 Style Transactions}
  \label{table:C5}
\end{table}

\section{Additional Tables and Visualizations}\label{Appendix D}

\begin{figure}
    \centering
    \includegraphics[width = \linewidth]{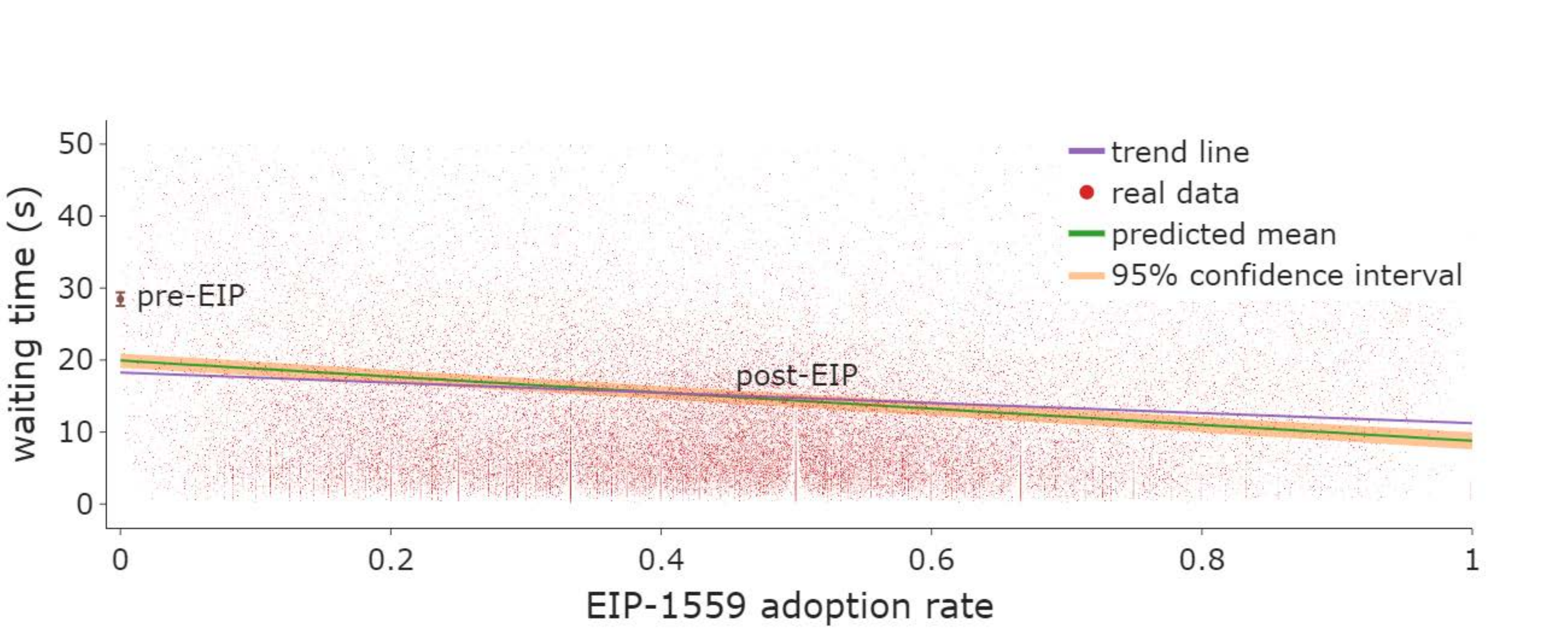}
    \begin{tablenotes}
    \footnotesize
    \item The figure visualizes Column (3) in~\cref{table: waiting time}. The line in the figure shows the predicted block median waiting time at different levels of the EIP-1559 adoption rate based on the linear regression outcome controlling for all other factors in a representative scenario (e.g., median gas price at 39 Gwei, price volatility at 0.005). As the adoption rate grows from 0 to 1, the predicted mean waiting time gradually decreases from 19.93 seconds to 8.79 seconds. Before the London hard fork, the average was approximately 28.5 seconds, shown as the red dot on the left side. The scatters are real data. The figure also includes another trend line that displays simple linear regression without controlling for any factors.
    \end{tablenotes}
    
    \caption{Simulated relationship between the waiting time and EIP-1559 adoption}
    \label{fig:cf_wt}
\end{figure}

\begin{figure}
    \centering
    \includegraphics[width = \linewidth]{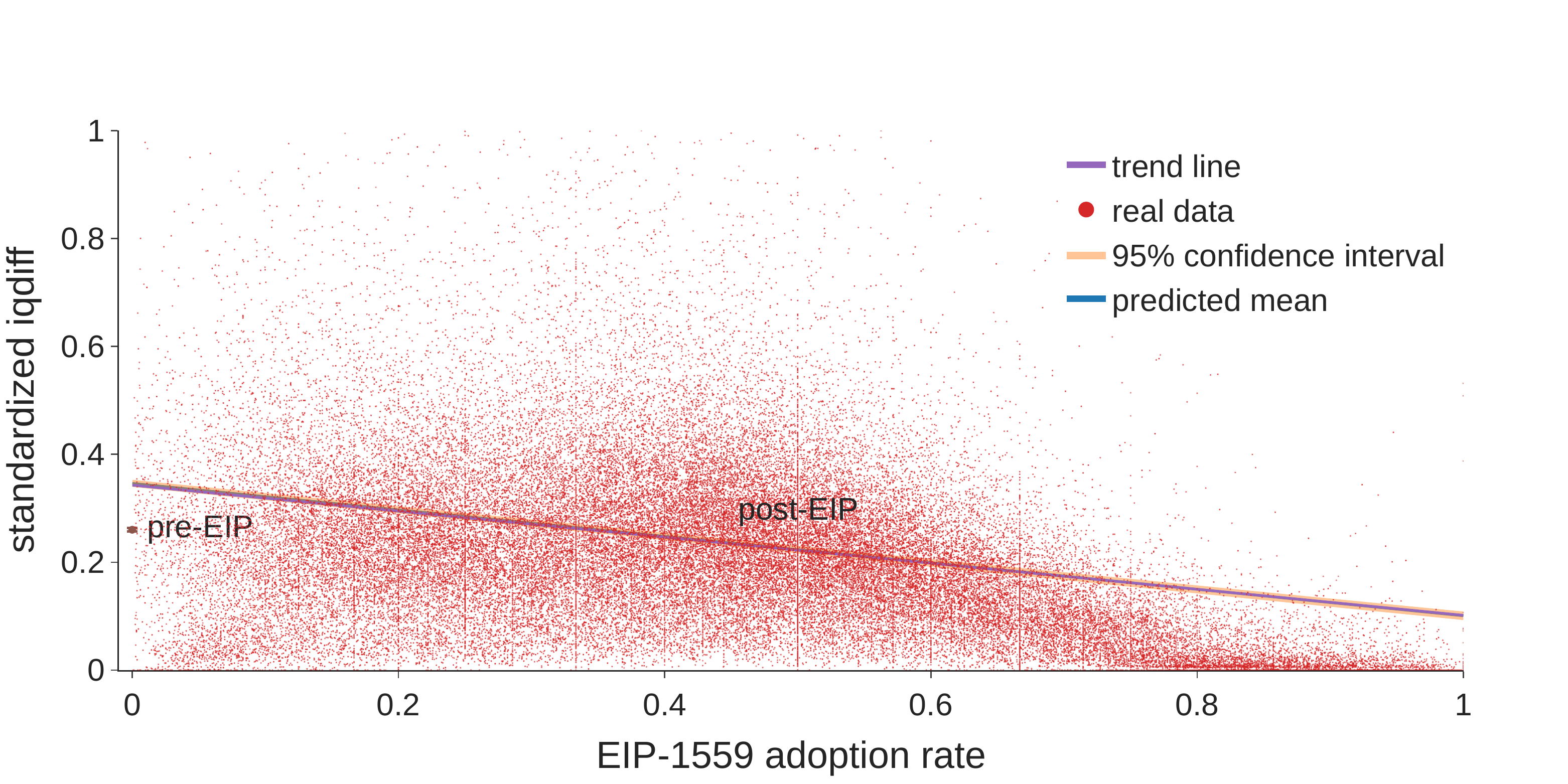}
    \begin{tablenotes}
    \footnotesize
    \item The figure visualizes Column (4) in Table 7. The line in the figure shows the predicted standardized IQR at different levels of the EIP-1559 adoption rate based on the linear regression outcome controlling for all other factors in a representative scenario (e.g., median gas price at 39 Gwei, price volatility at 0.005). As the adoption rate grows from 0 to 1, the predicted mean of the standardized interquartile difference gradually decreases from 0.34 to 0.09. For the period before the London hard fork, the average is approximately 0.27, shown as the red dot on the left side. The scatters are real data. The figure also includes another trend line that displays simple linear regression without controlling for any factors, although it largely coincides with the predicted line.
    \end{tablenotes}
    
    \caption{Simulated relationship between the standardized interquartile difference and EIP-1559 adoption}
    \label{fig:cf_iqdiff}
\end{figure}

\begin{figure}
  \centering
  \includegraphics[width = \linewidth]{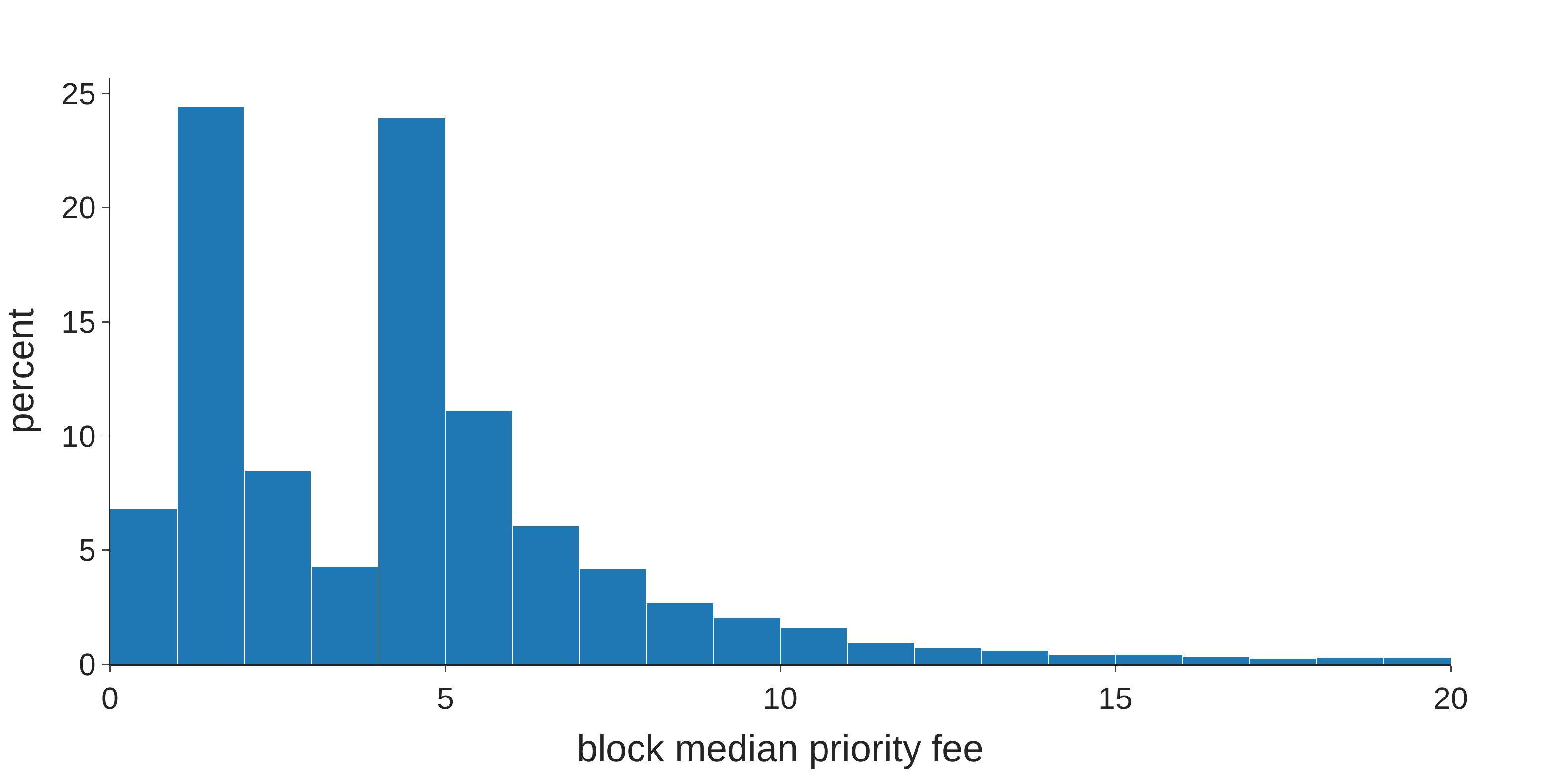}
  \caption{Distribution of priority fee bid}
  \label{fig:prdistr}
\end{figure}

Distribution of priority fees in~\cref{fig:prdistr} has two peaks at 2 Gwei and 5 Gwei, with a majority of median priority fee bids under 10 Gwei. 

\begin{figure}
  \centering
  \includegraphics[width = \linewidth]{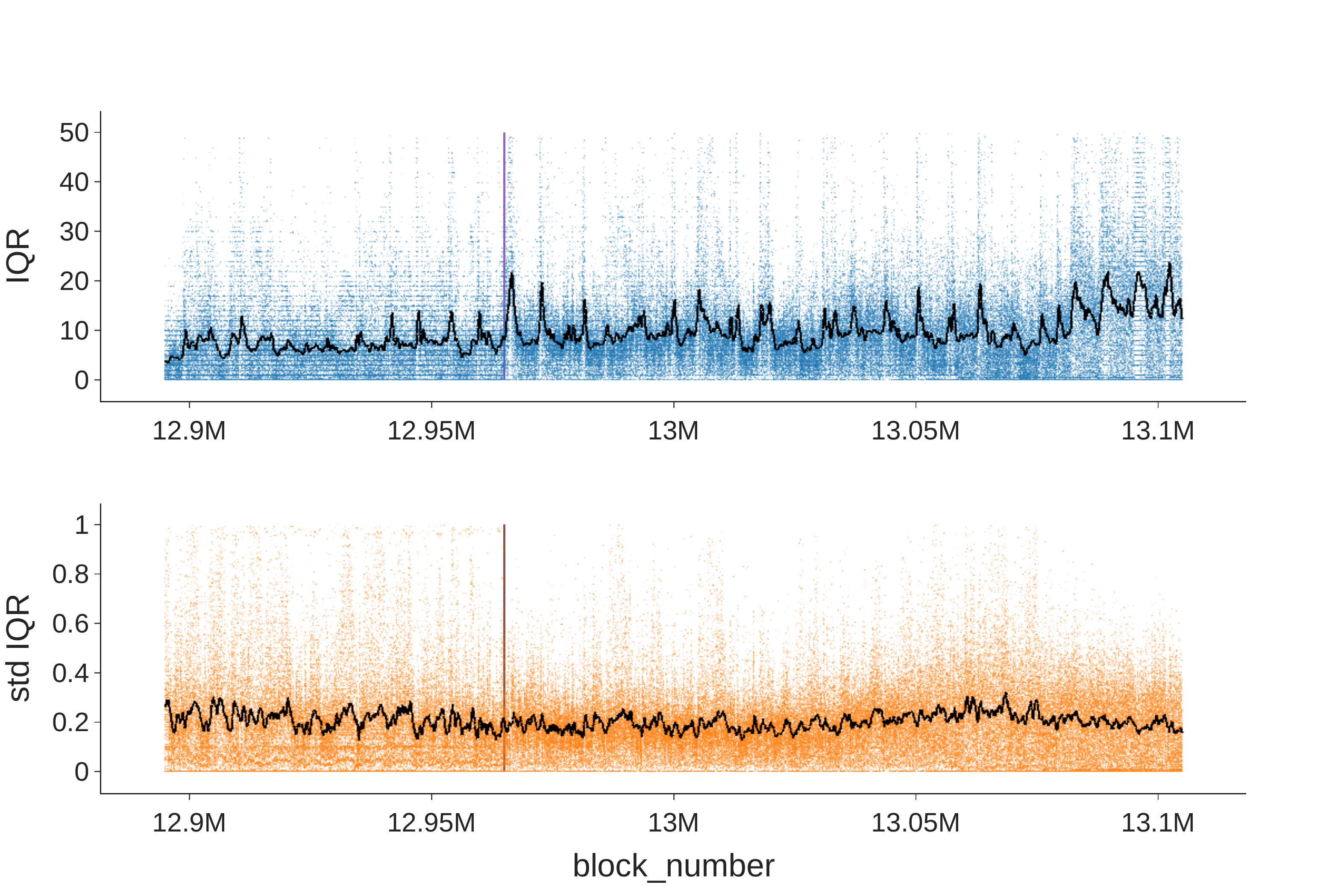}
  \caption{Time Series of intrablock gas price IQR and standardized IQR}
  \label{fig:IQR}
\end{figure}

\Cref{fig:IQR} shows the time series of IQR and standardized IQR, the former oscillating with base fee.

\begin{figure}
    \centering
  \includegraphics[width = \linewidth]{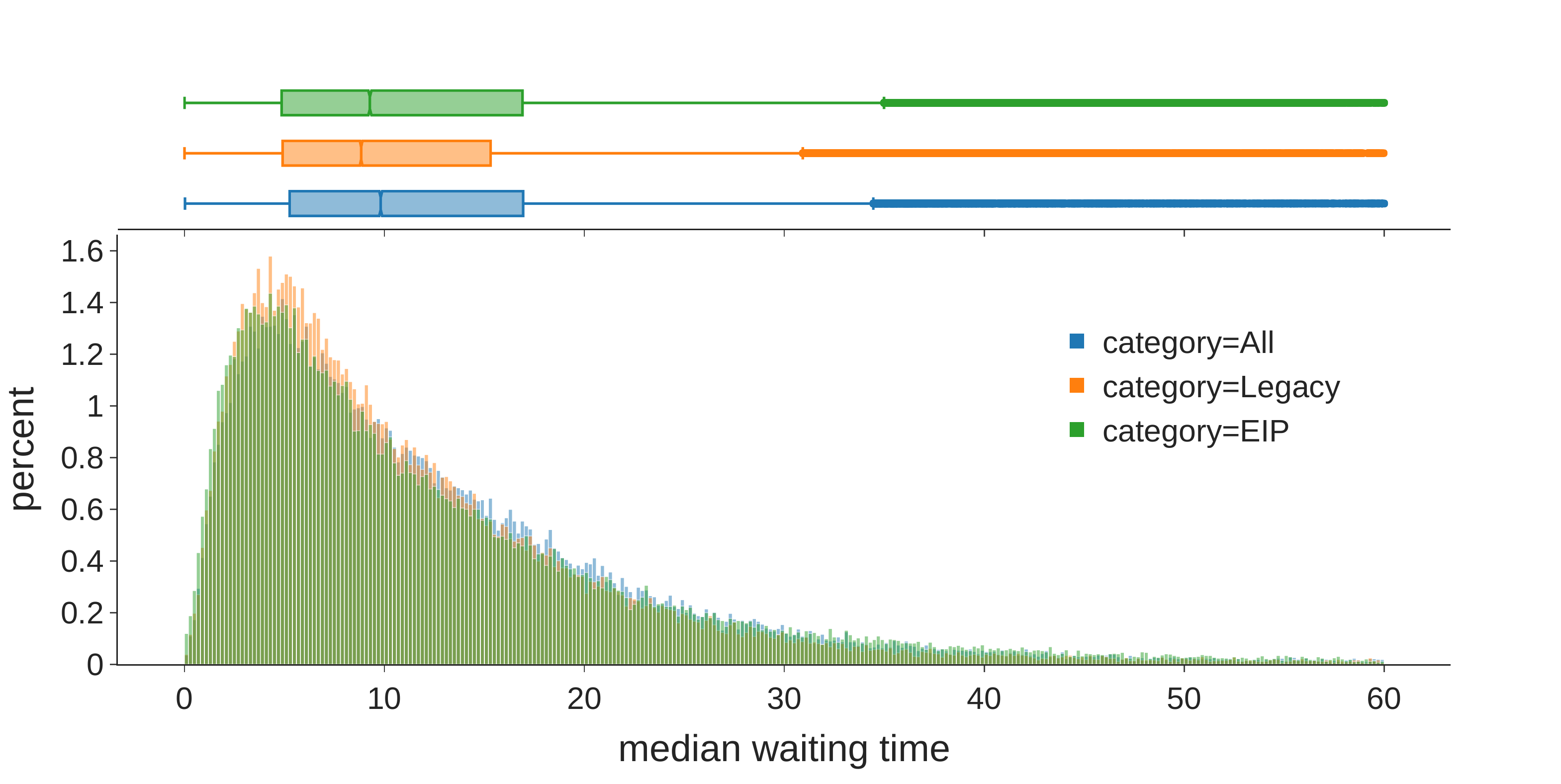}
  \caption{Median waiting time distribution by transaction type after London hard fork}
  \label{fig:wtdistrtype}
\end{figure}

\Cref{fig:wtdistrtype} shows that the waiting time distribution of legacy and EIP-1559 transactions after London hard fork, which are similar in general.

\Cref{Table 5.4.3} shows the percentage of network spikes for different time intervals and different thresholds.

\begin{table}[htbp]
    \centering
\small\setlength{\tabcolsep}{1.5pt}
{
\begin{tabular}{||c c c c c c c c c||} 
\multicolumn{3}{c}{\textbf{Pre London Hardfork}} \\ 
\hline
period$\backslash$threshold & 1.0 M & 1.2 M & 1.4 M & 1.6 M & 1.8 M & 2.0 M & 2.2 M & 2.4 M \\ 
\hline
20 s & 39.99 & 39.98 & 39.97 & 14.64 & 14.64 & 14.64 & 14.59 & 3.90\\
\hline
30 s & 35.29 & 34.40 & 34.39 & 14.67 & 14.66 & 5.07 & 4.92 & 4.92\\
\hline
40 s & 53.98 & 30.33 & 30.32 & 13.97 & 13.96 & 5.39 & 5.38 & 1.78\\
\hline
60 s & 42.22 & 41.90 & 24.26 & 12.37 & 5.57 & 2.26 & 2.23 & 0.83\\
\hline
90 s & 46.22 & 30.56 & 18.37 & 10.17 & 5.16 & 1.05 & 0.40 & 0.16\\
\hline
120 s & 49.20 & 35.10 & 14.41 & 8.25 & 2.30 & 0.52 & 0.24 & 0.03\\
\hline
\end{tabular}
\begin{tabular}{||c c c c c c c c c||} 
\multicolumn{3}{c}{\textbf{Post London Hardfork}} \\ 
\hline
period$\backslash$threshold & 1.0 M & 1.2 M & 1.4 M & 1.6 M & 1.8 M & 2.0 M & 2.2 M & 2.4 M \\ 
\hline
20 s & 39.11 & 31.88 & 25.66 & 15.12 & 10.38 & 7.27 & 5.00 & 3.38\\
\hline
30 s & 42.85 & 29.43 & 20.41 & 13.46 & 8.60 & 5.01 & 2.73 & 1.55\\
\hline
40 s & 45.57 & 31.45 & 20.46 & 11.41 & 6.35 & 3.35 & 1.76 & 0.86\\
\hline
60 s & 52.28 & 32.70 & 18.23 & 8.90 & 4.09 & 1.76 & 0.74 & 0.32\\
\hline
90 s & 57.77 & 33.83 & 16.27 & 6.54 & 2.35 & 0.80 & 0.30 & 0.13\\
\hline
120 s & 61.75 & 34.23 & 14.71 & 5.04 & 1.56 & 0.45 & 0.17 & 0.05\\
\hline
\end{tabular}
}

    \caption{Percentage of network spikes for different time intervals and different thresholds}
    \label{Table 5.4.3}
\end{table}

\begin{figure}
    \centering
  \includegraphics[width = \linewidth]{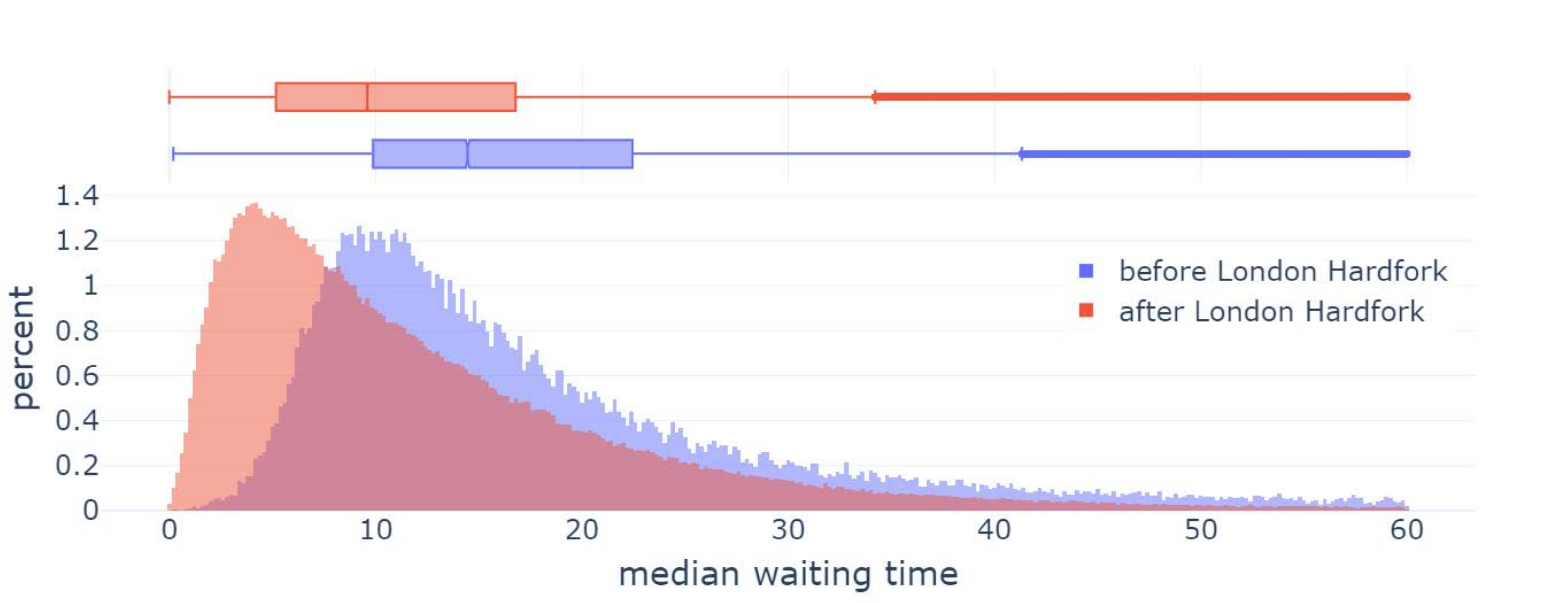}
  \caption{Distributions of median waiting time. It moved left after the London hard fork. Users experience a much lower transaction waiting time with EIP-1559. The result is robust even after we extend the time range for analysis.}
  \label{fig:wt_afterword}
\end{figure}

\Cref{fig:wt_afterword} represents the robustness check for the results on waiting time. 

\end{document}